\documentclass[%
twocolumn,
superscriptaddress,
 amsmath,amssymb,
 aps,prx
]{revtex4-2}

\usepackage{graphicx}
\usepackage{dcolumn}
\usepackage{bm}
\usepackage{color}
\usepackage{xspace}
\usepackage{multirow}
\usepackage{hyperref}
\usepackage{here}

\definecolor{green}{rgb}{0,0.6,0.1}

\newcommand{\NNO}{${\rm NdNiO}_{2}$\xspace}
\newcommand{\xx}{$d_{x^2-y^2}$\xspace}
\newcommand{\xy}{$d_{xy}$\xspace}
\newcommand{\zz}{$d_{3z^2-r^2}$\xspace}
\newcommand{\QE}{{\textsc{Quantum ESPRESSO}}\xspace}

\begin{document}

\preprint{APS/123-QED}

\title{Formation of 2D single-component correlated electron system and band engineering in the nickelate superconductor NdNiO$_2$}

\author{Yusuke Nomura}
\email{yusuke.nomura@riken.jp}
\author{Motoaki Hirayama}
\affiliation{ 
RIKEN Center for Emergent Matter Science, 2-1 Hirosawa, Wako, Saitama 351-0198, Japan 
}


\author{Terumasa Tadano}
\affiliation{
Research Center for Magnetic and Spintronic Materials, National Institute for Materials Science, Tsukuba 305-0047, Japan
}

\author{Yoshihide Yoshimoto}
\affiliation{
Department of Computer Science, The University of Tokyo,7-3-1 Hongo, Bunkyo-ku, Tokyo 113-0033
}

\author{Kazuma Nakamura}
\affiliation{
Quantum Physics Section, Kyushu Institute of Technology,1-1 Sensui-cho, Tobata, Kitakyushu, Fukuoka, 804-8550, Japan
}

\author{Ryotaro Arita}
\affiliation{ 
RIKEN Center for Emergent Matter Science, 2-1 Hirosawa, Wako, Saitama 351-0198, Japan 
}
\affiliation{
Department of Applied Physics, The University of Tokyo,7-3-1 Hongo, Bunkyo-ku, Tokyo 113-8656
}


\date{\today}

\begin{abstract}
Motivated by the recent experimental discovery of superconductivity in the infinite-layer nickelate Nd$_{0.8}$Sr$_{0.2}$NiO$_2$ 
[Li {\it et al.}, Nature {\bf 572}, 624 (2019)], we study how the correlated Ni 3\xx electrons in the NiO$_2$ layer interact with the electrons in the Nd layer. 
We show that three orbitals are necessary to represent the electronic structure around the Fermi level: Ni 3\xx, Nd 5\zz, and a bonding orbital made from an interstitial $s$ orbital in the Nd layer and the Nd 5\xy orbital.
By constructing a three-orbital model for these states, we find that the hybridization between the Ni 3\xx state and the states in the Nd layer is tiny. 
We also find that the metallic screening by the Nd layer is not so effective in that it reduces the Hubbard $U$ between the Ni 3\xx electrons just by 10--20\%. On the other hand, the electron-phonon coupling is not strong enough to mediate superconductivity of $T_{\rm c}\sim10$ K. These results indicate that \NNO hosts an almost isolated correlated 3\xx orbital system. We further study the possibility of realizing a more ideal single-orbital system in the Mott-Hubbard regime. We find that the Fermi pockets formed by the Nd-layer states dramatically shrink when the hybridization between the interstitial $s$ state and Nd 5\xy state becomes small. By an extensive materials search, we find that the Fermi pockets almost disappear in NaNd$_2$NiO$_4$ and NaCa$_2$NiO$_3$. 
\end{abstract}


\maketitle


\section{Introduction}

Since the discovery of high transition temperature ($T_{\rm c}$) superconductivity in the cuprates~\cite{Bednorz_1986}, a wide variety of two-dimensional correlated materials
including layered transition metal oxides, pnictides, chalcogenides, some heavy fermion materials, and organic conductors 
have provided a unique playground for unconventional superconductivity~\cite{Stewart_2017}.  Among them, the high $T_{\rm c}$ cuprates are of great interest due to their remarkably simple electronic structure: The Fermi surface consists of only one two-dimensional band. Thus one may expect that if just one correlated orbital makes a two dimensional Fermi surface, 
high $T_{\rm c}$ superconductivity will emerge~\cite{Arita_1999,Arita_2000,Monthoux_1999,Monthoux_2001,Sakakibara_2010, Sakakibara_2012,Sakakibara_2012b}.

However, it has been a great theoretical and experimental challenge to find an analogue of the cuprates~\cite{Norman_2016}. One possible candidate is vanadates, for which we can think of designing a $d^1$-analogue of the cuprates~\cite{Pickett_1989,Imai_2005,Matsuno_2005,Arita_2007}. However, it is not an easy task to separate the 
two-dimensional $3d_{xy}$-band from the $3d_{yz}$ and $3d_{zx}$ bands. While nickelates can be an interesting $d^7$-analogue of the cuprates,
we need to consider a special 
heterostructure to remove the $3d_{3z^2-r^2}$ band from the Fermi level~\cite{Chaloupka_2008,Hansmann_2009,Hansmann_2010,Han_2011}. More recently, Sr$_2$IrO$_4$ and Ba$_2$IrO$_4$ are attracting broad interest as a possible 5$d$-analogue of the cuprates~\cite{Kim_2008,Kim_2009,Watanabe_2010,Watanabe_2013}.
While it has been recently reported that doped Sr$_2$IrO$_4$ exhibits spectroscopic signatures suggesting a $d$-wave superconducting gap~\cite{Kim_2015,Yan_2015}, a zero-resistance state has not yet been observed. 
It should be noted that the system cannot be regarded as an ideal single-orbital system in that not only the $j_{\rm eff}=1/2$ state but also $j_{\rm eff}=3/2$ states are involved with the low-energy physics~\cite{Arita_2012,Martins_2012}.
Therefore, ``orbital distillation"~\cite{Sakakibara_2012} in two-dimensional correlated systems has been a key issue for realizing an analogue of the cuprates.

Recently, Li {\it et al.} discovered that the infinite-layer nickelate NdNiO$_2$ exhibits superconductivity when Sr is doped~\cite{Li_2019}. It is interesting to note that NdNiO$_2$ is isostructural to the infinite-layer copper oxide CaCuO$_2$ which becomes a superconductor below $T_{\rm c}=110$ K by hole doping~\cite{Azuma_1992}.
In this compound, Ni has square planar oxygen coordination  
and is expected to have a $d^9$ configuration. 
Due to the absence of the apical oxygen, the crystal field in \NNO is totally different from the octahedral ligand field. Notably, the energy level of the $3d_{3z^2-r^2}$ orbital becomes much lower than that of the $3d_{x^2-y^2}$ orbital, even lower than the orbitals labeled $t_{2g}$ in the octahedral environment~\cite{Botana_arXiv}. 
In this situation, only the $3d_{x^2-y^2}$ band among the five 3$d$ bands intersects the Fermi level. 
While this is good news for designing an analogue of the cuprates, the previous calculations based on density functional theory (DFT) pointed out that there are two complications~\cite{Lee_2004,Botana_arXiv,Sakakibara_arXiv,Jiang_arXiv,Wu_arXiv}. One is that electrons in the Nd layer make additional Fermi pockets. The other is that the charge transfer gap between the transition-metal $3d$ states and O $2p$ states is larger than that in the cuprates.


In this paper, we focus on the first problem, i.e., the role of the ``additional'' Nd 5$d$ Fermi pockets which are absent in the cuprates. Especially, we discuss the following issues of critical importance: i) What is the minimal model to describe these additional degrees of freedom in the Nd layer? ii) Does the hybridization between the Ni 3$d_{x^2-y^2}$ band and these additional bands invalidate the effective single-band 3$d_{x^2-y^2}$ model?
iii) Does the metallic screening from the additional Fermi pockets significantly weaken the effective Hubbard interaction between the Ni 3$d_{x^2-y^2}$ electrons and help the conventional phonon mechanism?
iv) Can we switch the presence/absence of the additional Fermi pockets?

For these issues, we show that:
i) We need two orbitals to describe the additional low-energy bands: the Nd 5\zz orbital and interstitial $s$ orbital (or the Nd 5\xy orbital~\cite{Sakakibara_arXiv}). 
ii) In the three-orbital model for the Ni 3\xx orbitals and two orbitals in the Nd layer, the hybridization between the Ni and Nd layer is weak; thus, the Ni 3\xx electrons are almost isolated. 
iii) The metallic screening by the Nd layer is not significant, leaving the Ni 3\xx electrons strongly correlated. On the other hand, the electron-phonon coupling is not sufficiently strong to mediate superconductivity of $T_{\rm c}\sim10$ K. 
vi) By controlling the hybridization between the interstitial state and the Nd 5$d_{xy}$ orbital, we can change the size of the additional Fermi pockets. More specifically, we show that the additional Fermi pockets almost disappear in NaNd$_2$NiO$_4$ and NaCa$_2$NiO$_3$.
While the cuprates belong to the charge-transfer type in the Zaanen-Sawatzky-Allen classification~\cite{Zaanen_1985}, these nickelate compounds will provide an interesting playground to investigate the possibility of high $T_{\rm c}$ superconductivity in the Mott-Hubbard regime.


The structure of the paper is as follows. After describing the detail of the methods in Sec.~\ref{sec_methods}, in Sec.~\ref{sec_DFT_band}, we first show the band dispersion and the Bloch wave functions at some specific $\bm{k}$ points. We will see that the Fermi surface consists of three bands: The Ni 3$d_{x^2-y^2}$ band, Nd 5$d_{3z^2-r^2}$ band, and that formed by a bonding state between interstitial and Nd 5\xy orbitals in the Nd layer. 
To exclude the possibility of phonon-mediated superconductivity, we calculate the phonon frequencies and electron-phonon coupling in Sec.~\ref{sec_phonon}. We show that the electron-phonon coupling constant is much smaller than 0.5, so that phonons cannot mediate superconductivity of $T_{\rm c}\sim 10$ K. 
To consider the possibility of unconventional superconductivity, we then derive an effective low-energy model of NdNiO$_2$ in Sec.~\ref{sec_effective_model}.  
By looking at the tight-binding parameters in the Hamiltonian, and plotting the projected density of states (PDOSs), we show that the hybridization between the Ni 3$d_{x^2-y^2}$ state and the Nd-layer states is small. In Sec.~\ref{sec_cRPA}, we perform a calculation based on the constrained random phase approximation (cRPA)~\cite{Aryasetiawan_2004}, and estimate the interaction parameters in the three-orbital model and single-orbital model. By comparing the values of the Hubbard $U$, we show that the screening effect of the Nd-layer on the 3\xx orbital is not so strong. 
In Sec.~\ref{sec_enhance_Tc}, we discuss how we can change the band dispersion and realize an ideal single-component two-dimensional correlated electron system. We show that the hybridization between the interstitial state and the Nd $5d_{xy}$ state is the key parameter for the materials design.


\section{Methods}
\label{sec_methods}

Here, we describe the calculation methods and conditions used in the following sections. 

\paragraph{ Secs.~\ref{sec_DFT_band}, \ref{sec_effective_model}, and \ref{sec_cRPA}}
The DFT band structure calculations for \NNO are performed using \QE~\cite{QE-2017} with
the experimental lattice parameters of bulk \NNO: $a = 3.9208$ \AA\xspace and $c = 3.2810$ \AA~\cite{Hayward_2003}.    
We use Perdew-Burke-Ernzerhof (PBE)~\cite{Perdew_1996} norm-conserving pseudopotentials generated by the code ONCVPSP (Optimized Norm-Conserving Vanderbilt PSeudopotential)~\cite{Hamann_2013}, which are obtained from the PseudoDojo~\cite{Setten_2018}.
We use 11$\times$11$\times$11 $\bm{k}$ points for sampling in the first Brillouin zone. 
The energy cutoff is set to be 100 Ry for the wave functions, and 400 Ry for the charge density. 
After the band structure calculations, the maximally localized Wannier functions~\cite{Marzari_1997,Souza_2001} are constructed 
using RESPACK~\cite{RESPACK_URL}\footnote{Algorithms and applications of RESPACK can be found in Refs.~\cite{Nakamura_2016,Nakamura_2009,Nakamura_2008,Nohara_2009,Fujiwara_2003}.}. 
The three-orbital model consisting of Ni 3\xx, Nd 5\zz and interstitial $s$ orbitals is constructed using
the outer energy window of  [$-2$ eV : 10 eV] and the inner energy window of [0 eV : 1.5 eV]. 
The Ni 3\xx single orbital model is constructed using 
the outer energy window of  [$-1.15$ eV : 2.05 eV] without setting the inner energy window.
The interaction parameters are calculated with RESPACK~\cite{RESPACK_URL}
using the constrained random-phase approximation (cRPA) method~\cite{Aryasetiawan_2004}, in which we employ the scheme in Ref.~\onlinecite{Sasioglu_2011} for the band disentanglement.  
The energy cutoff for the dielectric function is set to be 20 Ry. 
The total number of bands used in the calculation of the polarization is 80,
which includes the unoccupied states up to $\sim 55$ eV with respect to the Fermi level.

To evaluate the effects due to the difference in DFT codes and pseudopotentials, we also derive the low-energy effective models using xTAPP~\cite{Yamauchi_1996}. Although the basic calculation conditions are the same as those of \QE, the xTAPP calculations employ the pseudopotentials generated as follows: Ni pseudopotential is generated in a slightly ionic semicore $(3s)^{2}(3p)^{6}(3d)^{9}$ configuration by employing the cutoff radii of $r_{3s}=r_{3p}=r_{3d}=0.8$ bohr. The Nd pseudopotential is constructed under the configuration of 
$(5s)^2(5p)^6(5d)^1(6s)^2$, while that of the core electrons is (Kr)$(4f)^3$; the $4f$ electrons are frozen and excluded from the pseudopotential. 
The cutoff radius for the local potential is 1.3 bohr, and those for $s$, $p$, $d$ orbitals of the pseudopotential are 1.3, 1.9, and 1.9 bohr, respectively. The $5s$ and $6s$ orbitals share the same cutoff radius. We apply the partial-core correction with a cutoff radius of 1.1 bohr.
To stabilize the band calculation under the fairly hard pseudopotential of Ni, the xTAPP calculations adopt the energy cutoffs of 196 Ry and 784 Ry for wave function and the charge density, respectively. With this condition, almost perfect agreement between the \QE and xTAPP results is obtained for the low-energy band structure and the Wannier function.

\paragraph{ Sec.~\ref{sec_phonon}}

Phonon band structure, electron-phonon coupling constants, and the Eliashberg function of \NNO are calculated based on density functional perturbation theory (DFPT)~\cite{Baroni_2001}, as implemented in \QE~\cite{QE-2017}. We use PBE pseudopotentials obtained from the pslibrary 1.0.0~\cite{DALCORSO_2014}. The $4f$ orbitals of Nd is treated as a frozen core. The kinetic energy cutoff is set to 75 Ry for the wave function and 600 Ry for the charge density. The self-consistent field calculation is conducted with 16$\times$16$\times$16 $\bm{k}$ mesh, and DFPT calculation is performed with 8$\times$8$\times$8 $\bm{q}$ mesh. Before performing DFPT calculation, the lattice constants are optimized until the stress convergence criterion of $|\sigma| < 0.5$ kbar is reached. The optimized values are $a=3.893$ \AA{} and $c=3.267$ \AA{}, which reproduce the reported experimental values within 1\% error.

\paragraph{ Sec.~\ref{sec_enhance_Tc} } 

We use Vienna \textit{ab initio} simulation package (\textsc{VASP})~\cite{kresse_1996} for optimizing the lattice parameters of LiNd$_{2}$NiO$_{4}$, NaNd$_{2}$NiO$_{4}$, and NaCa$_{2}$NdO$_{3}$.
The wave-function cutoff energy of 50 Ry is employed along with the PBE PAW potentials~\cite{Blochl_1994,Kresse_1999} labeled as \texttt{Li\_sv}, \texttt{Na\_pv}, \texttt{Ca\_sv}, \texttt{Nd\_3}, \texttt{Ni\_pv}, and \texttt{O}.
The 12$\times$12$\times$8 $\bm{k}$ mesh is employed for the Brillouin zone sampling.
For the calculation of the electronic band structures, we use the \textit{ab initio} code OpenMX~\cite{Ozaki_2003}.
We employ the same $\bm{k}$ mesh as that of the VASP calculation.
We employ the valence orbital sets Li8.0-$s3p3d2$, Na9.0-$s3p2d1$,  Ca9.0-$s4p3d2$, Nd8.0\_OC-$s2p2d2f1$, Ni6.0H-$s4p3d2f1$, and O-5.0$s3p3d2$.
The energy cutoff for the numerical integration is set to 150 Ry.

\section{Formation of single-orbital correlated system in ${\rm NdNiO}_2$}

\subsection{Band structure}  
\label{sec_DFT_band}

\begin{figure}[tb]
\vspace{0cm}
\begin{center}
\includegraphics[width=0.48\textwidth, clip]{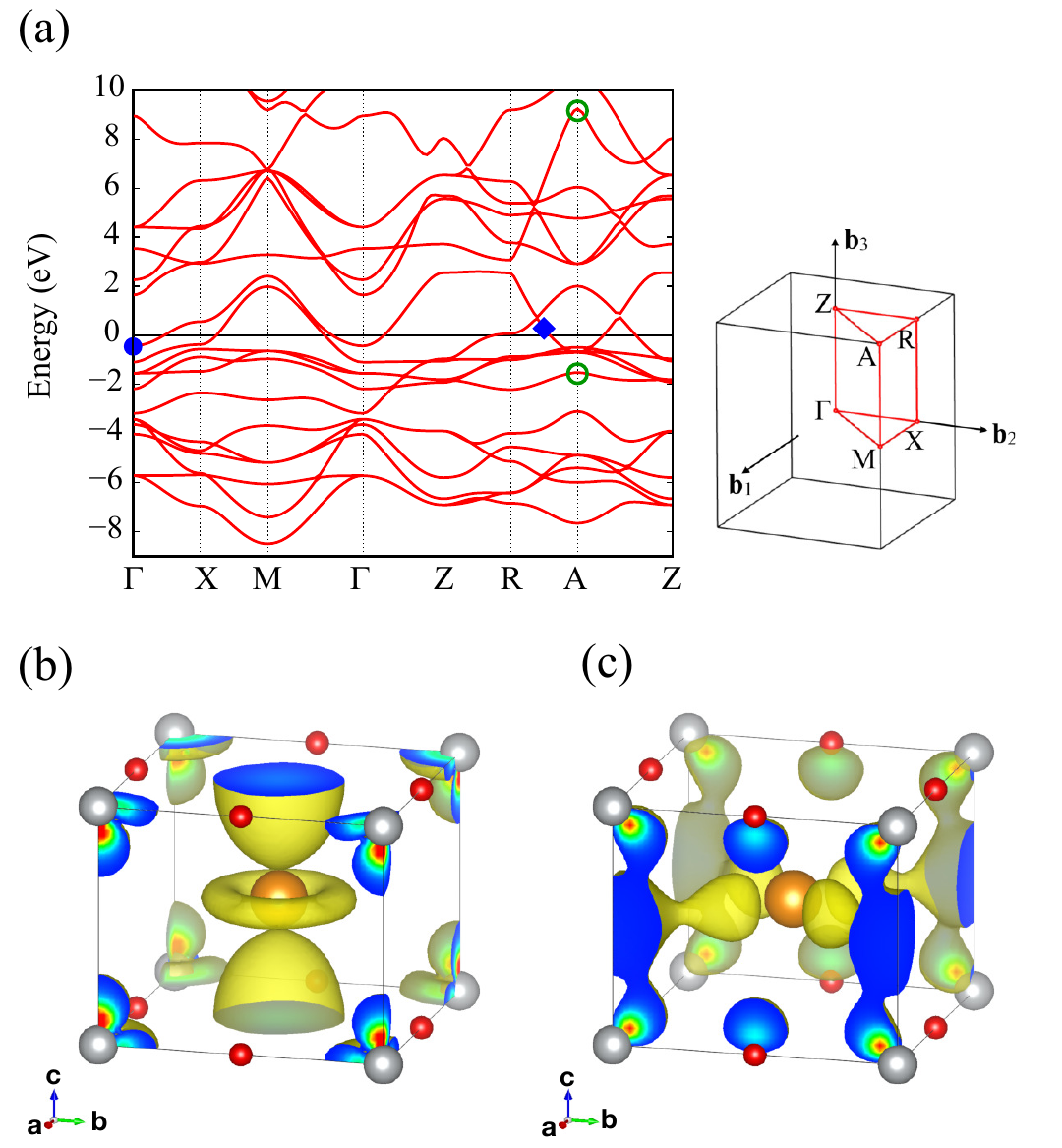}
\caption{
(a) DFT band structure for NdNiO$_2$.  
The Nd $4f$ orbitals are assumed to be frozen in the core.
In \NNO, the interstitial $s$ and Nd 5\xy orbitals 
form bonding and antibonding states on the $k_z = \pi/c$ plane. 
The green open circles at A point indicate
the band bottom of the bonding band (below the Fermi level)
and the band top of the antibonding band 
(above the Fermi level). 
Electron density of the Kohn-Sham states specified by (b) the blue dot at $\Gamma$ point and (c) the blue diamond at ($\pi/a$, $\pi/2a$, $\pi/c$) point 
(drawn by VESTA~\cite{Momma_2011}). 
The positions of the atoms in the crystal coordinate are follows: 
Ni (0, 0, 0), Nd (1/2, 1/2, 1/2), O (1/2, 0, 0), O (0, 0, 1/2). 
}
\label{Fig_DFT_band}
\end{center}
\end{figure}

Figure ~\ref{Fig_DFT_band}(a) shows the DFT band structure for NdNiO$_2$.
In the calculations, the Nd $4f$ orbitals are assumed to be frozen in the core so that dispersionless $4f$ bands do not appear in the band structure.

We see that the Ni 3$d$ bands (located from $\sim -3$ eV to $ \sim 2 $ eV)
and the O $2p$ bands (located from $\sim -9$ eV to $\sim -3$ eV) are well separated, which is in high contrast with the cuprate counterpart CaCuO$_2$~\cite{Botana_arXiv}.
Because of the absence of the apical oxygen, the energy level of the Ni 3$d_{3z^2-r^2}$ orbital becomes far below the Fermi level. 
Therefore, as far as the Ni $3d$ orbitals are concerned, only the 3$d_{x^2-y^2}$ orbitals form the Fermi surface. 

As already pointed out by previous studies~\cite{Lee_2004,Botana_arXiv}, on top of the Fermi surfaces originated from the Ni 3\xx orbital, 
there are additional Fermi pockets around $\Gamma$ and A points. 
Around $\Gamma$ point, as can be confirmed by the electron density of the Kohn-Sham orbital [Figure~\ref{Fig_DFT_band}(b)], the Fermi pocket originates mainly from the Nd $5d_{3z^2 - r^2}$ orbital~\cite{Lee_2004,Botana_arXiv}.  
On the other hand, the orbital character of the Fermi pocket around A point shows an interesting feature [Figure ~\ref{Fig_DFT_band}(c)]. 
The electron density has a large weight at the interstitial region around (0, 0, 1/2) site, which corresponds to the site of the apical oxygen in NdNiO$_3$.
This interstitial state is stabilized because (0, 0, 1/2) site is surrounded by cations. 
Therefore, the NdNiO$_2$ system has a feature similar to electrides where electrons at interstitial regions play a role of anions~\cite{Matsuishi03,Hirayama18e}.
The interstitial orbitals form a bonding state with the Nd 5$d_{xy}$ orbital. 
The band bottom of the bonding orbital is located at A point [green open circle below the Fermi level in Fig.~\ref{Fig_DFT_band}].
The energy difference between the bonding and antibonding states at A point is more than 10 eV. 
As we will discuss in Sec.~\ref{sec_enhance_Tc}, we can realize a single-component correlated system by controlling the energy dispersion of the interstitial bonding state.

\subsection{Phonons} 
\label{sec_phonon}

Because \NNO has a different Fermi surface topology from the cuprates, there might be a possibility that the superconducting mechanism is different. 
One of the most fundamental questions would be whether the observed superconductivity can be explained by the phonon mechanism or not. 
To investigate this point, we performed phonon calculations and estimated the strength of the electron-phonon coupling. 

\begin{figure}[tb]
  \centering
  \includegraphics[width=0.49\textwidth, clip]{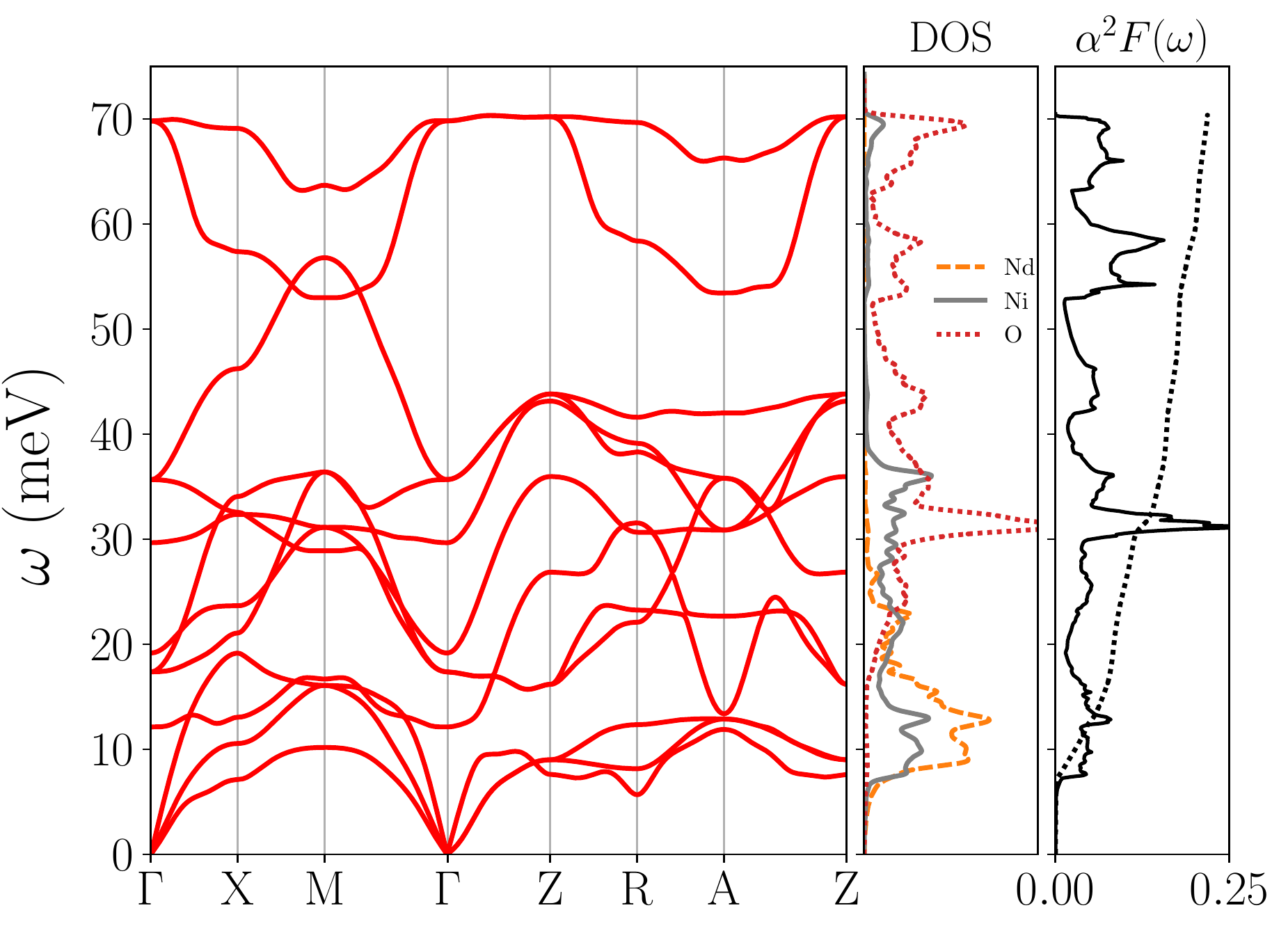}
  \caption{
  Phonon band structure (left panel), atom-projected phonon DOS (middle), and the Eliashberg function $a^2F(\omega)$ (right) of \NNO. The dotted line in the right panel shows the accumulated value of $\lambda(\omega)$ defined as $\lambda(\omega) = 2\int_{0}^{\omega}d\omega' \alpha^{2}F(\omega')/\omega'$.}
  \label{Fig_pnonon_band}
\end{figure}

  Figure \ref{Fig_pnonon_band} shows the calculated phonon and electron-phonon properties of \NNO. The phonon band structure shows no imaginary modes, confirming the dynamical stability of bulk \NNO. 
  To evaluate the Brillouin zone-averaged electron-phonon coupling strength $\lambda$, we first calculated the coupling strength for each phonon mode $\nu$ at momentum $\bm{q}$ defined as 
  \begin{equation*}
  \lambda_{\bm{q}\nu} = \frac{2}{N(\epsilon_{\mathrm{F}})N_k\omega_{\bm{q}\nu}}\sum_{mn\bm{k}}|g_{mn,\nu}(\bm{k},\bm{q})|^{2}\delta(\xi_{m\bm{q}})\delta(\xi_{n\bm{k}+\bm{q}}).
  \end{equation*}
  Here, $\omega_{\bm{q}\nu}$ is the phonon frequency, $N(\epsilon_{\mathrm{F}})$ is the DOS at the Fermi level, $N_{k}$ is the number of $\bm{k}$ points, $g_{mn,\nu}(\bm{k},\bm{q})$ is the electron-phonon matrix element, and $\xi_{m\bm{k}}=\epsilon_{m\bm{k}}-\epsilon_{\mathrm{F}}$ is the energy of the Kohn-Sham orbital $m$ at momentum $\bm{k}$ relative to the Fermi level. The double delta function of $\lambda_{\bm{q}\nu}$ was evaluated by the Gaussian smearing method with different smearing width $\sigma$. We then calculated the Eliashberg function $\alpha^{2}F(\omega)$ and $\lambda$ as $\alpha^{2}F(\omega)=\frac{1}{2N_q}\sum_{\bm{q}\nu}\lambda_{\bm{q}\nu}\omega_{\bm{q}\nu}\delta(\omega-\omega_{\bm{q}\nu})$ and $\lambda = 2\int_0^{\infty} d\omega \frac{\alpha^{2}F(\omega)}{\omega}$, respectively. The results for $\alpha^{2}F(\omega)$ and the accumulated value of $\lambda$ is shown in the right panel of Fig.~\ref{Fig_pnonon_band}.

  The electron-phonon interaction evaluated with $\sigma=0.04$ Ry is $\lambda = 0.22$. Although the $\lambda$ value changes with the smearing width as shown in Table \ref{table:el-ph}, we confirmed that it does not reach 0.5 with reasonably small values of $\sigma$. 
  We note that the $\lambda$ values of \NNO are as small as those calculated for the cuprate superconductors YBa$_{2}$Cu$_{3}$O$_{7}$ ($\lambda=0.27$, Ref.~\onlinecite{Bohnen_2003}) and La$_{2-x}$Sr$_{x}$CuO$_{4}$ ($\lambda=$ 0.14--0.22, Ref.~\onlinecite{Giustino_2008}). The logarithmic average of phonon frequencies and the $T_{\mathrm{c}}$ values of \NNO are also summarized in Table \ref{table:el-ph}. The $T_{\mathrm{c}}$ values, obtained using the Allen-Dynes formula~\cite{Allen_1975} with $\mu^{*}=0.1$, are too small to account for the experimental results of $T_{\mathrm{c}} = $ 9--15 K. Therefore, we can rule out the electron-phonon interaction as the exclusive origin of the superconductivity observed in the doped \NNO.

\begin{table}[htb]
\caption{
Electron-phonon interaction $\lambda$ and the logarithmic average of phonon frequencies calculated for \NNO with different width of the Gaussian smearing. The $T_{\mathrm{c}}$ values are evaluated using the Allen-Dynes formula with $\mu^{*}=0.1$.
}
\label{table:el-ph}
\centering 
\vspace{0.1cm}
\begin{ruledtabular}
\begin{tabular}{cccc}
Smearing width (Ry) &  $\lambda$ & $\omega_{\ln}$ (K) & $T_{\mathrm{c}}$ (K) \\ \hline
0.04 & 0.22 & 283 & 0.00 \\
0.06 & 0.28 & 258 & 0.06 \\
0.08 & 0.32 & 249 & 0.24 \\
\end{tabular} 
\end{ruledtabular}
\end{table}

\subsection{Minimal tight-binding Hamiltonian} 
\label{sec_effective_model}

\begin{figure*}[htbp]
\vspace{0cm}
\begin{center}
\includegraphics[width=0.99\textwidth, clip]{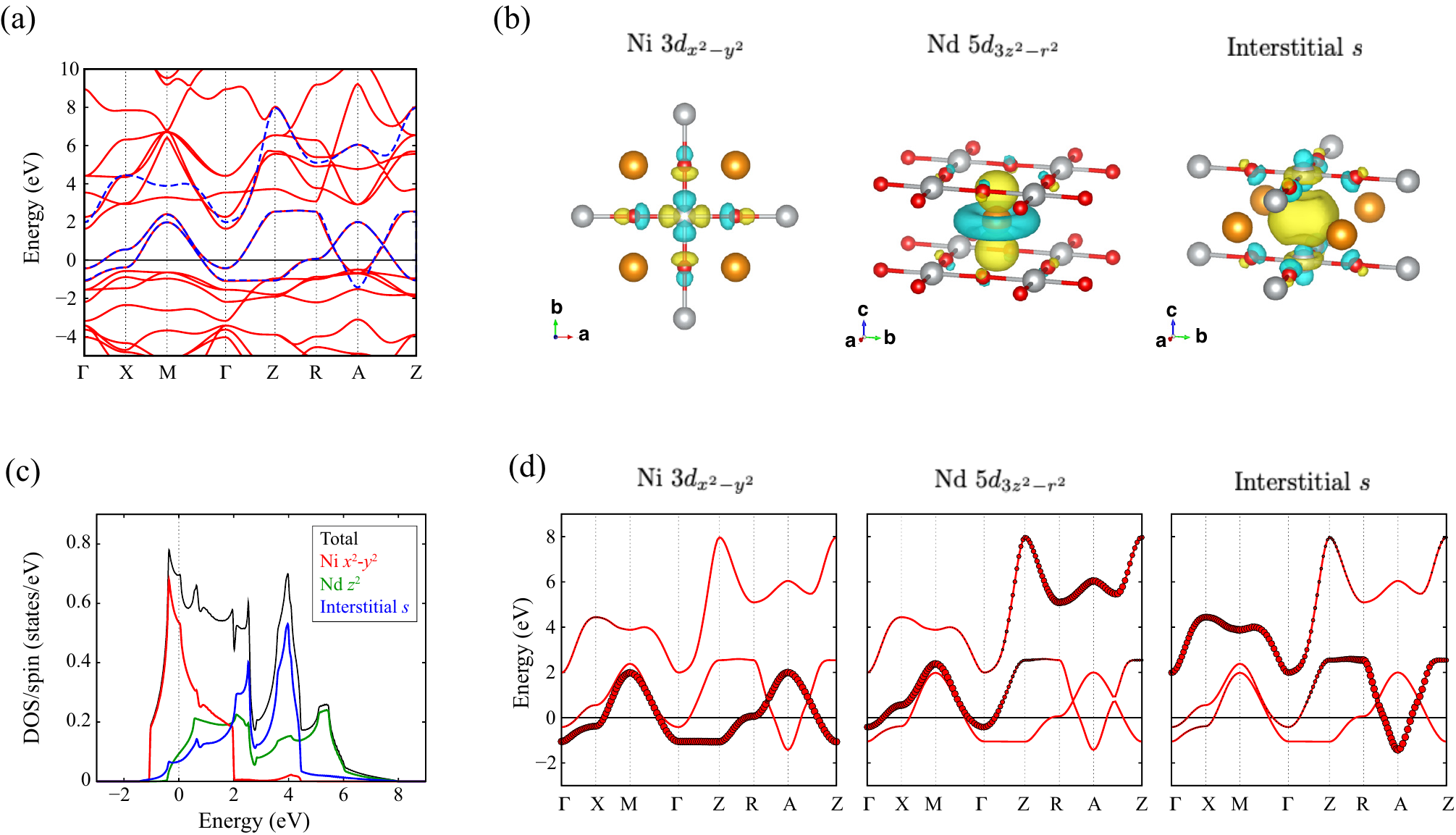}
\caption{
(a) Dispersion of the three-orbital tight binding model (blue dashed curves).  Red solid curves show the DFT band structure. 
(b) Isosurfaces (yellow: positive, light blue: negative) of constructed maximally localized Wannier functions (drawn by VESTA~\cite{Momma_2011}). 
(c) Projected density of states (PDOS) and (d) fat bands of the three-orbital tight binding model. In the fat bands, the size of the symbols is proportional to the weight of the respective orbital components.
}
\label{Fig_Wannier_band}
\end{center}
\end{figure*}

In Section \ref{sec_phonon}, we have seen that the electron-phonon coupling is not strong enough to explain the experimental transition temperature. 
Next, we consider the possibility of whether the Ni 3$d$ electron systems can be a playground for unconventional superconductivity.
To analyze the electron correlation effect, 
it is essential to derive realistic tight-binding Hamiltonian for the low-energy electrons near the Fermi level~\cite{Imada_2010,Kotliar_2006}.
Here, as a minimal tight-binding model to reproduce the band structure around the Fermi level, based on the orbital character analysis in Figs.~\ref{Fig_DFT_band}(b) and (c), we propose a three-orbital model consisting of the Ni 3\xx orbital, Nd 5\zz orbital, and interstitial $s$ orbital centered at (0, 0, 1/2) site.

Figure~\ref{Fig_Wannier_band}(a) shows a comparison between the original DFT band structure (red solid curves) and the band dispersion obtained from the three-orbital tight-binding model (blue dashed curves). 
We see that the band structure near the Fermi level is accurately reproduced. 
The panel (b) visualizes the constructed maximally localized Wannier functions,
in which we see that the Wannier functions have characters of Ni 3\xx, Nd 5\zz, and interstitial $s$, respectively. 
The Ni 3\xx Wannier orbital has tails of O $2p$ orbitals and form the antibonding orbital. 
Similarly, the interstitial $s$ orbital has an antibonding character with the Ni 3\zz orbital.
It should be noted that the lobes of the Nd 5\xy orbital with $(\pi/a, \pi/a)$ phase on the $x$-$y$ plane
(then the phases of the lobes from Nd 5\xy orbitals
at (0, 0, 1/2), (1, 0, 1/2), (0, 1, 1/2), (1, 1, 1/2) sites are aligned)
participate in the interstitial $s$ Wannier orbital
and form the bonding state. 
This is the reason why the additional Fermi pocket around
A point [${\bf k} = (\pi/a,\pi/a,\pi/c)$] can be described by the interstitial-$s$-centered bonding orbital  
or Nd-5\xy-centered bonding orbital.
See Appendix \ref{App_3orb_dxy} for the construction of another three-orbital model using the Nd-5\xy-centered bonding orbital~\cite{Sakakibara_arXiv}.

We show the calculated density of states (DOS) for the three-orbital model and the decomposition into the three orbitals [Ni 3\xx (red), Nd 5\zz (green), and the interstitial $s$ (blue)] in Fig.~\ref{Fig_Wannier_band}(c), 
and the fat bands (bands weighted by the orbital components) in Fig.~\ref{Fig_Wannier_band}(d). 
Notably, the hybridization between the Ni 3\xx orbital and the other orbitals is negligibly small and the Ni 3\xx orbital form an isolated single band. 
From the symmetry consideration, we can show that the transfer between the neighboring Ni 3\xx and Nd 5\zz orbitals is zero. 
The same is true for the transfer between the neighboring Ni 3\xx and interstitial $s$ orbitals. 
These facts and a large difference in the onsite level between the Ni 3\xx and 
Nd 5\zz and the interstitial $s$ orbitals (about 2.6 eV and 2.4 eV, respectively) 
help the isolation of the Ni 3\xx orbital. 

We see that at the Fermi level, the majority of the DOS is from the Ni 3\xx orbital, but both the Nd 5\zz and the interstitial $s$ orbitals also have finite DOS.
The additional Fermi surface around $\Gamma$ point is formed by the Nd 5\zz orbital, whereas that around A point is formed by the interstitial $s$ orbital, in accord with the orbital character analysis in Figs.~\ref{Fig_DFT_band}(b) and (c). 
The absence of the apical oxygen makes the lattice constant along the $z$ direction small, resulting in the large bandwidth of the Nd-layer orbitals. 
Because of the finite occupation of the Nd 5\zz and the interstitial $s$ orbitals, the 
occupation of the Ni 3\xx orbital deviates from half-filling.
The calculated occupation is about 0.87, corresponding to 13 \% hole doping. This number is consistent with an estimate in the previous study~\cite{Botana_arXiv}. 

The absence of apical oxygen makes the Ni 3\xx band highly two-dimensional.
It can be confirmed by the shape of the PDOS of Ni 3\xx and the negligible dispersion along the $z$ direction.


\subsection{Correlation strength} 
\label{sec_cRPA}

The analysis in section~\ref{sec_effective_model} leads to the conclusion that the low-energy electronic structure of \NNO is well characterized by the highly-two dimensional Ni 3\xx band and two independent Nd-layer bands. 
Then, an important question is whether Ni 3\xx bands are strongly correlated or not. To answer this question, we estimate the strength of the effective Coulomb repulsion using the cRPA method~\cite{Aryasetiawan_2004}.

First, we calculate the interaction parameters for the three-orbital model. 
Table~\ref{param-3orb} shows our calculated effective interaction parameters of the three-orbital model. 
The Ni 3\xx onsite interaction $U$ is about 3.1 eV, which is comparable to the bandwidth of the Ni 3\xx band (about 3 eV). 
Compared to $U$ for the Ni 3\xx orbital, $U$ for the Nd 5\zz and the interstitial $s$ orbitals are smaller, reflecting their larger spatial spread. 

%

\begin{table}[tb] 
\caption{
Calculated onsite interaction parameters $U$ 
in the three-orbital model. 
Unit is given in eV.
{\sc Quantum Espresso} and xTAPP give consistent results. 
} 
\vspace{0.2cm} 
\centering 
\begin{tabular}{l@{\ \ \ }c@{\ \ \ }c} \hline \hline \\ [-8pt]
                 & {\sc Quantum Espresso} & xTAPP  \\ [3pt] \hline \\ [-5pt]
$U_{{\rm Ni}(3d_{x^2-y^2})}$ & 3.109 & 3.129                 \\ [4pt] 
$U_{{\rm Nd}(5dz^2)}$        & 2.108 & 2.133                 \\ [4pt] 
$U_{{\rm Interstitial}(s)}$  & 1.075 & 1.121                 \\ [4pt]
\hline \hline
\end{tabular} 
\label{param-3orb} 
\end{table}

If we take account of the metallic screening by the Nd-layer orbitals, the effective interaction parameter for the Ni 3\xx orbital will be reduced.  
To investigate this screening effect, we calculate 
the interaction parameter for the Ni 3\xx single-orbital model by including the Nd-layer screening effect. 
Table~\ref{param-1orb} shows the main Coulomb interaction parameters of the single-band model.  
We also show the hopping parameters in the table. 
What is interesting is that even when the Nd-layer metallic screening effect is incorporated, the Ni 3\xx orbital has an onsite interaction of about 2.6 eV, and the reduction of $U$ is less than 20 \%. 
We note that if the Ni 3\xx intraband metallic screening effect is further taken into account, $U$ is drastically reduced down to about 1.2 eV. 
Therefore, the effect of the metallic screening by electrons in the Nd layer on the Ni 3\xx onsite interaction is weak.

The mechanism behind the weak screening effect is summarized by the following three points: 
i) The electron density of the Nd-layer orbitals is spatially separated from the NiO$_2$ plane. Therefore, the electron clouds of the Nd-layer orbitals cannot directly screen the Coulomb interaction in the NiO$_2$ plane. 
ii) The Nd-layer polarization is not strong because the DOS of the Nd-layer orbitals is small at the Fermi level.
iii) The orbital off-diagonal polarization involving the Ni 3\xx orbital and the Nd-layer orbitals is small because of the tiny hybridization as discussed in the previous section. 
More detailed discussion on the role of the Nd-layer screening can be found in Appendix~\ref{App_7_and_5_cRPA}. 

Therefore, the Ni 3\xx orbital feels strong onsite Coulomb repulsion. 
The nearest transfer $t$ is nearly $-0.37$ eV, and thus the ratio $|U/t|$ amounts to around 7, whose value is comparable to that estimated for the cuprates~\cite{Tadano_2019,Hirayama_2018,Hirayama_2019,Werner_2015,Jang_2016,Nilsson_2019}. 
Therefore, we conclude that the Ni 3\xx electrons in \NNO 
forms strongly-correlated two-dimensional system.

As for other parameters in the table, $V_a$ and $V_c$ represent nearest neighbor interactions along the $a$ and $c$ axes, respectively, which are sufficiently smaller than $U$. $t'$ and $t''$ are the next-nearest and third-nearest transfer integrals, respectively. In particular, frustration degree of freedom in the square lattice $t'/t$ is nearly $-0.25$, which is comparable to that of cuprate~\cite{Botana_arXiv,Pavarini_2001}


\begin{table}[tb] 
\caption{Calculated interaction parameters in the single-orbital model. $U$ is onsite interaction and $V_a$ and $V_c$ represent nearest neighbor interactions along the $a$ and $c$ axes, respectively. $t$, $t'$, $t''$ are the nearest, next-nearest, and third-nearest transfer integrals, respectively. Unit is given in eV.
{\sc Quantum Espresso} and xTAPP give consistent results. 
}
\vspace{0.2cm} 
\centering 
\begin{tabular}{c@{\ \ \ }c@{\ \ \ }c} \hline \hline \\ [-8pt]
                 & {\sc Quantum Espresso} & xTAPP  \\ [3pt] \hline \\ [-5pt]
 $U$    & \ 2.608 & \  2.578     \\ [4pt] 
 $V_a$  & \ 0.218 & \ 0.223     \\ [4pt] 
 $V_c$  & \ 0.143 & \  0.154     \\ [4pt]
 $t$    & $-0.370$ & $-0.375$  \\ [4pt] 
 $t'$   & \ 0.092 & \ 0.091     \\ [4pt] 
 $t''$  & \ $-0.045$ & \ $-0.044$  \\ [4pt] 
 $|U/t|$  & \ 7.052 & \  6.874     \\ [4pt] 
 $t'/t$ & $-0.250$ & $-0.243$     \\ [4pt] 
\hline \hline
\end{tabular} 
\label{param-1orb} 
\end{table}






\section{Band engineering for realizing pure Mott-Hubbard type single-band system}

\label{sec_enhance_Tc}

\begin{figure}[tb]
\vspace{0cm}
\begin{center}
\includegraphics[width=0.48\textwidth, clip]{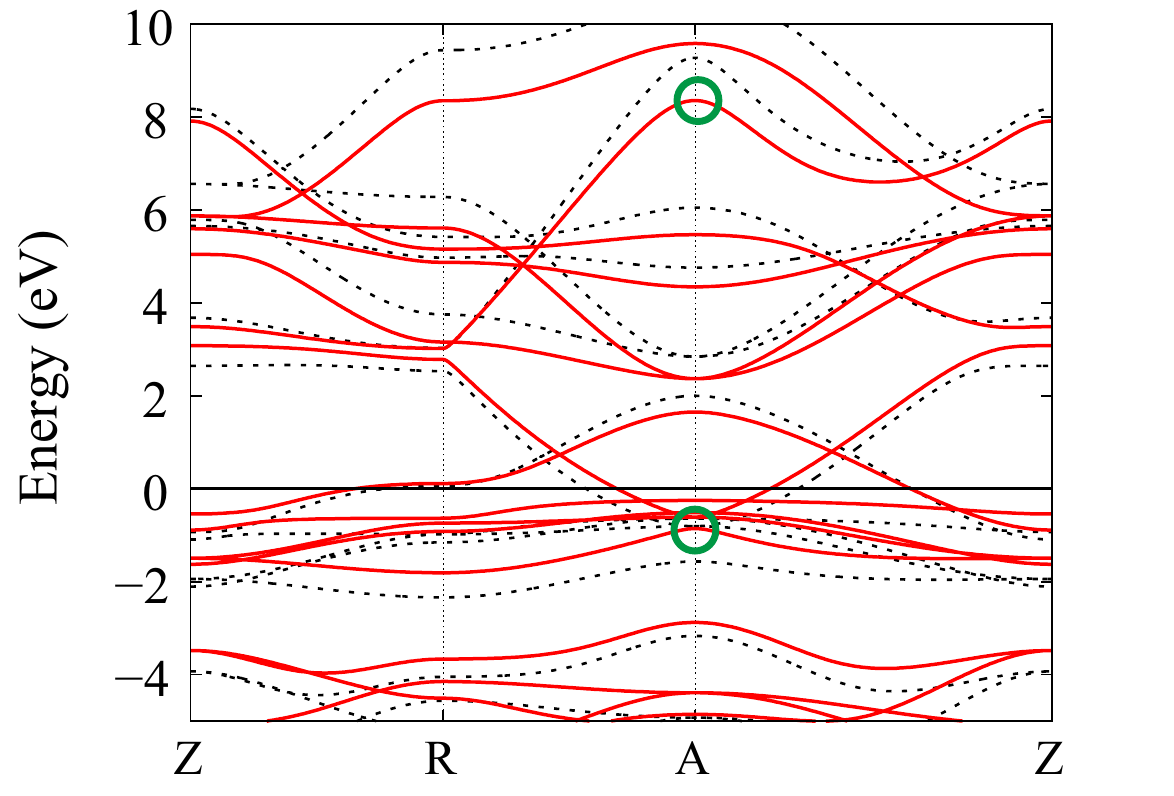}
\caption{
DFT band structure of NdNiO$_2$.
The black dotted and red solid curves are calculated with the original lattice constant $a$ and $1.05 \times a$, respectively. 
The green open circles indicate
the band bottom of the bonding band between the interstitial $s$ and Nd 5\xy orbital 
(below the Fermi level)
and the band top of the antibonding band 
(above the Fermi level). 
}
\label{NdNiO2_c1.05_band}
\end{center}
\end{figure}

\begin{figure*}[htbp]
\vspace{0cm}
\begin{center}
\includegraphics[width=0.99\textwidth, clip]{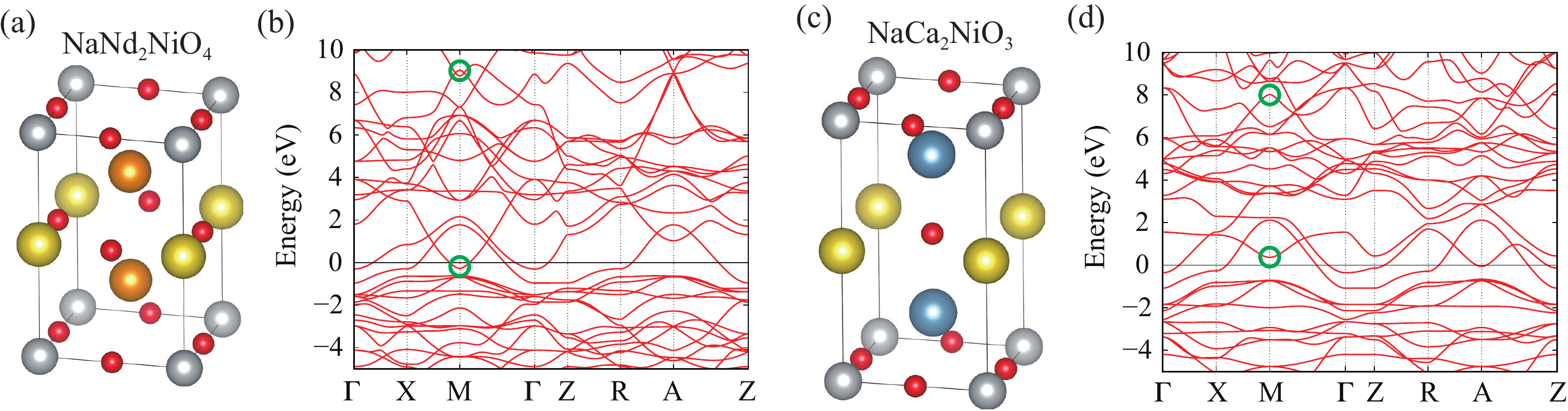}
\caption{
(a) Crystal structure and (b) DFT band structure of NaNd$_2$NiO$_4$.
(c) Crystal structure and 
(d) DFT band structure of NaCa$_2$NiO$_3$.
The green open circles indicate
the band bottom of the bonding band between the interstitial $s$ and cation \xy orbital 
(below the Fermi level)
and the band top of the antibonding band 
(above the Fermi level). 
}
\label{NaNd2NiO4_NaCa2NiO3_band}
\end{center}
\end{figure*}

\begin{table*}[ptb] 
\caption{Energy of the interstitial $s$ and the cation $d_{xy}$ orbitals ($\Delta E_{s}$ and $\Delta E_{d_{xy}}$, respectively) for NdNiO$_2$, LiNd$_2$NiO$_4$, NaNd$_2$NiO$_4$, and NaCa$_2$NiO$_3$.
We also show the energies for the bottom of the bonding band $E_b$ and the top of the antibonding band $E_a$.
$\Delta E_{sd}$ and $\Delta E_{ba}$ are the energy difference between $s$ and $d_{xy}$ states, and bonding and antibonding band, respectively.
Here, $d_{xy}$ indicates the Nd $5d_{xy}$ orbital for NdNiO$_2$, LiNd$_2$NiO$_4$, and NaNd$_2$NiO$_4$, 
and the Ca $3d_{xy}$ orbital for NaCa$_2$NiO$_3$.
$E_s$ and $E_{d_{xy}}$ are estimated, respectively, from R and Z points in \NNO, 
and from X and $\Gamma$ points in the other materials.
In these ${\bf k}$ points, the hybridization between the interstitial $s$ or $d_{xy}$ orbitals and the other orbitals become small.
$E_b$ and $E_a$ are estimated from A point in NdNiO$_2$,
and M point in the other materials.
The unit for length is \AA.
The energy unit is eV.
} 
\vspace{0.2cm}
\begin{tabular}{@{\ \ \ }c@{\ \ \ }|@{\ \ \ }c@{\ \ \ }|@{\ \ \ }c@{\ \ \ }c@{\ \ \ }c@{\ \ \ }c@{\ \ \ }|@{\ \ \ }c@{\ \ \ }c@{\ \ \ }}
\hline \hline \\ [-8pt]   
 & $a$  & E$_s$ & E$_{d_{xy}}$ & E$_b$ & E$_a$ & $\Delta E_{sd}$ & $\Delta E_{ba}$  \\ [+1pt]
\hline \\ [-8pt] 
NdNiO$_2$ & 3.921 (exp.) & 2.529 & 3.676 & $-$1.567 & 9.270 & 1.147 & 10.837 \\  [+1pt]
\hline \\ [-8pt] 
LiNd$_2$NiO$_4$ & 3.908 & 3.116 & 4.212 & $-$1.014 & 9.545 & 1.096 & 10.559 \\
NaNd$_2$NiO$_4$ & 4.056 & 3.739 & 3.908 & $-$0.271 & 9.027 & 0.169 & 9.298 \\ 
NaCa$_2$NiO$_3$ & 3.918 & 1.421 & 4.777 & \ \ 0.368 & 8.018 & 3.356 & 7.650 \\ 
\hline \hline 
\end{tabular}
\label{Ene_sxy} 
\end{table*} 

So far, we have shown that 
the electronic structure of 
\NNO can be accurately described by a single-orbital strongly-correlated Ni 3\xx band coupled weakly to the Nd-layer electrons. 
Considering the large charge transfer energy, 
\NNO would belong to the Mott-Hubbard type~\cite{Jiang_arXiv} rather than the charge-transfer type in the Zaanen-Sawatzky-Allen classification scheme~\cite{Zaanen_1985}. 
Therefore, the nickelate would be an interesting playground for investigating superconductivity in the Mott-Hubbard type single-band system, which will promote the understanding of the high $T_{\rm c}$ superconductivity in the cuprates.

Here, we propose a way to perform ``band engineering" to realize a more ideal two-dimensional Mott-Hubbard type single-band system, 
in which almost only the Ni 3\xx band intersects the Fermi surface.
For this purpose, we need to eliminate the Fermi pocket around $\Gamma$ and A points originated from the Nd-layer band. 
 Because the Fermi pocket around the $\Gamma$ point is already small, we hereafter focus on the Fermi pocket around the A point. 

As we have already discussed in Sec.\ref{sec_DFT_band}, 
the band dispersion of the bonding state between the interstitial $s$ and Nd 5\xy orbitals takes its minimum
at A point in $\bm{k}$ space \footnote{We note that on ${k_z} = 0$ plane, 
the interstitial $s$ and Nd 5\xy orbitals can also hybridize to Ni 3\zz orbital, whereas on ${k_z} = \pi/c$ plane, the hybridization is zero by symmetry. 
This makes a difference between 
${k_z} = 0$ and ${k_z} = \pi/c$ planes, where 
the former has no Fermi pocket around M point while 
the latter has the Fermi pocket around A point.
}. 
The energy of the band bottom is located well below the Fermi level ($\sim -1.5$ eV).
The energy of the antibonding state at A point is located around 9 eV, and the energy difference is larger than 10 eV (see Fig.~\ref{Fig_DFT_band}). 

In order to eliminate the Fermi pocket from the bonding orbital, there are basically two strategies: 
i)  reduce the bandwidth and make the energy difference between the bonding and antibonding states smaller. 
ii) raise the energy center of the bonding-antibonding gap. 
The former is related to the strength of the hopping between the interstitial $s$ and Nd 5\xy orbitals, whereas the latter is related to their onsite energy levels. 
Because the hopping amplitude is easier control, we follow the former strategy.

To reduce the hopping amplitude between the interstitial $s$ and Nd 5\xy orbitals,
the simplest way is to increase the in-plane lattice constant $a$.
By increasing $a$ by 5 \%, we indeed see that the energy difference between the bonding and antibonding states at the A point becomes smaller 
and that the band bottom of the bonding state becomes closer to the Fermi level 
(see Fig.\ref{NdNiO2_c1.05_band}). 
However, the effect is not strong enough to eliminate the Fermi surface.

Alternatively, let us introduce a new block anion layer, then we can reduce the hopping amplitude without applying the tensile strain. 
Here, the unit cell consists of 
two anion layers (NiO$_2$ and new anion layer) and two cation layers (such as La and Nd). 
It should be noted that the mirror inversion symmetry with respect to the cation layers is broken. 
Then, the height of the center of the interstitial $s$ and cation 5\xy orbitals becomes different. 
Because of the bending of the hopping path, the hopping amplitude is expected to decrease. 

We have verified several different possibilities for the new block layers, such as LiO$_2$, NaO$_2$, LiO, and NaO. 
In the case of LiO and NaO, to satisfy the charge neutrality, we change the cation layer from Nd to Sr or Ca. 
Among them, we find that 
NaNd$_2$NiO$_4$ and NaCa$_2$NiO$_3$, whose structures are shown in Figs.~\ref{NaNd2NiO4_NaCa2NiO3_band}(a), and (c), 
are potential candidates to realize an ideal two-dimensional Mott-Hubbard type correlated system. 
The band structures for NaNd$_2$NiO$_4$ and NaCa$_2$NiO$_3$ are shown in Figs.~\ref{NaNd2NiO4_NaCa2NiO3_band}(b), and (d), respectively.
In these materials, because there exist two anion layers, 
M point corresponds to A point in \NNO. 
Remarkably, in both materials, the Fermi pockets around M point almost disappear.

Indeed in these two materials, as shown in Table~\ref{Ene_sxy},
the energy difference between the bonding and antibonding states $\Delta E_{ba}$ (9.298 and 7.650 eV) 
is significantly smaller than that of \NNO (10.837 eV).  
Especially in NaCa$_2$NiO$_3$, even though the in-plane lattice constant $a$ is smaller than that of \NNO, 
we see a significant decrease in $\Delta E_{ba}$, which can be ascribed to the bending of the hopping path on the cation layer. 
The increase in the energy difference between the interstitial $s$ and Nd 5\xy orbitals (see $\Delta E_{sd}$) makes the hybridization smaller, and this is also helpful in reducing $\Delta E_{ba}$. 
Finally, let us mention that the effect is material dependent. 
For example, in LiNd$_2$NiO$_4$, the decrease in $\Delta E_{ba}$ is not so drastic and the Fermi pocket survives (not shown). 
However, we see that the decrease in $\Delta E_{ba}$ is a general tendency, therefore, we believe that the present results provide a useful guildeline for future material synthesis.

\section{Summary } 

In this paper, we studied the role of the states in the Nd layer in the newly discovered nickelate superconductor, \NNO. While only the 3\xx band among the five 3$d$ bands intersects the Fermi level, we need at least two states to represent the Fermi pockets formed by the states in the Nd layer: One is the Nd 5\zz state, and the other is an interstitial $s$ state, which makes the situation similar to electrides. In the three-orbital model for these states, the hybridization between the Nd-layer states and Ni 3\xx state is small. We performed a calculation based on the cRPA to estimate the interaction parameters. By comparing the results for the three-orbital model and single-orbital model for the Ni 3\xx orbital, we found that the screening effect of the Nd-layer states is not so effective, reducing the Hubbard $U$ between the Ni 3\xx electrons just by 10--20\%. On the other hand, the electron-phonon coupling constant is not larger than 0.5, so that the phonon mechanism is unlikely. Given the results indicating that \NNO hosts an almost ideal single-component correlated 3\xx electrons, we further study whether we can eliminate the Fermi pockets formed by the Nd-layer states. We found that these Fermi pockets dramatically shrink if the hybridization between the interstitial $s$ state and the Nd 5\xy states is reduced. By an extensive materials search, we found that the Fermi pockets almost disappear in NaNd$_2$NiO$_4$ and NaCa$_2$NiO$_3$. These nickelates will provide an interesting playground to investigate the possibility of high $T_{\rm c}$ superconductivity in the Mott-Hubbard regime.

\begin{acknowledgments}
We acknowledge the financial support of JSPS Kakenhi Grant No. 16H06345 (YN, MH, TT, YY, KN, and RA), No. 17K14336 (YN), No. 18H01158 (YN), No. 16K05452 (KN), No. 17H03393 (KN), No. 17H03379 (KN), No. 19K03673 (KN), and No. 19H05825 (RA).

\end{acknowledgments}

\appendix


\section{ Screening process for the onsite interaction of Ni 3\xx orbital } 
\label{App_7_and_5_cRPA}

\begin{table*}[ptb] 
\caption{
Comparison of cRPA onsite interactions among seven-orbital, five-orbital, three-orbital, and single-orbital models. 
Unit is given in eV.
{\sc Quantum Espresso} and xTAPP give consistent results. 
} 
\vspace{0.2cm} 
\centering 
\begin{tabular}{l@{\ \ }c@{\ \ }c@{\ \ }c@{\ \ }c@{\ \ }c@{\ \ }c@{\ \ }c@{\ \ }c@{\ \ }c} 
\hline \hline \\ [-8pt]
 & \multicolumn{4}{c}{{\sc Quantum Espresso}} & & \multicolumn{4}{c}{xTAPP} \\ \hline \\ [-8pt]  
 & 7 orbital & 5 orbital & 3 orbital & 1 orbital & 
 & 7 orbital & 5 orbital & 3 orbital & 1 orbital 
 \\ \hline \\ [-8pt] 
$U_{{\rm Ni}(3d_{x^2-y^2})}$ & 
 5.112 & 2.933 & 3.109 & 2.608 & & 5.144 & 2.841 & 3.129 &  2.578  \\ [4pt] 
$U_{{\rm Ni}(3d_{z^2})}$ & 
 6.824 & 2.649 & -     & -     & & 6.757 & 2.793 & -     & -     \\ [4pt] 
$U_{{\rm Ni}(3d_{xy})}$ & 
 5.501 & 3.294 & -     & -     & & 5.488 & 3.171 & -     & -     \\ [4pt] 
$U_{{\rm Ni}(3d_{yz})}$ & 
 5.841 & 3.096 & -     & -     & & 5.837 & 3.099 & -     & -     \\ [4pt] 
$U_{{\rm Ni}(3d_{zx})}$ & 
 5.841 & 3.096 & -     & -     & & 5.837 & 3.099 & -     & -     \\ [4pt] 
$U_{{\rm Nd}(5dz^2)}$  & 
 2.972 & -     & 2.108 & -     & & 2.950 & -     & 2.134 & -      \\ [4pt] 
$U_{{\rm Interstitial}(s)}$ & 
 2.491 & -     & 1.075 & -     & & 2.453 & -     & 1.121 & -      \\ [4pt]
\hline \hline
\end{tabular} 
\label{param-7orb} 
\end{table*}

\begin{figure*}[ptb] 
\centering
\vspace{0.5cm} 
\includegraphics[width=14cm, clip]{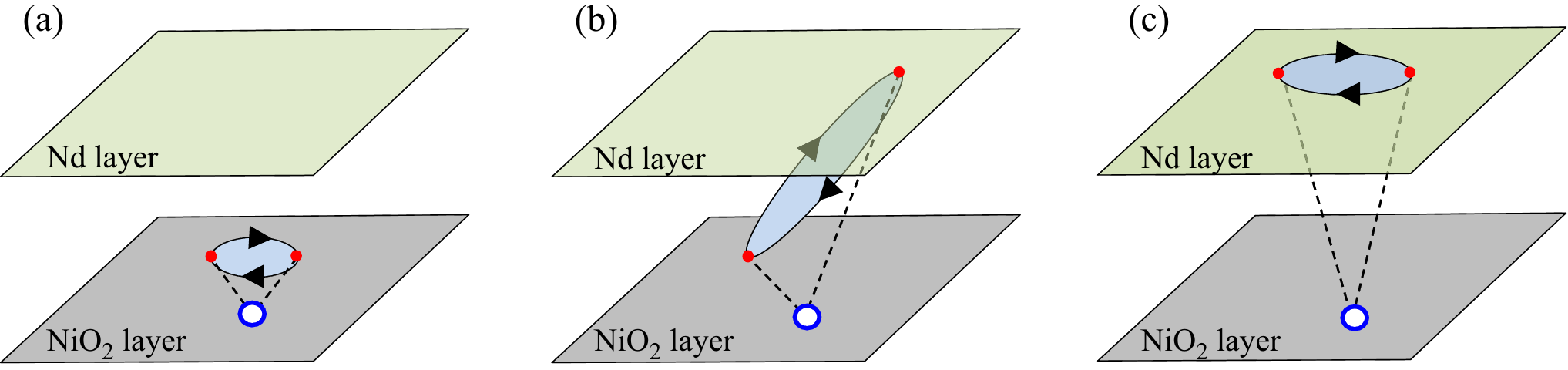}
\caption{
Schematic diagram for 
screening processes for 
the 3\xx
orbital
at a Ni site (blue circles). Upper-green and lower-gray layers represent Nd and NiO$_2$ layers, respectively. Dashed lines are Coulomb interactions, and solid oval lines with arrows describe the polarization in which the red dots represent the polarization points. In the figure, an RPA diagram (symbolically for the second-order one) for the effective onsite interaction is shown, 
and three 
types of
screening processes are depicted: 
(a) Screening process via the intralayer polarization in the NiO$_2$ layer. (b) Screening process via the interlayer polarization. (c) Screening process via the intralayer polarization in Nd layer.
The red points and the blue circles in (a) and (b) may share the same Ni site, although they are depicted at difference positions. 
} 
\label{layer-screening}
\end{figure*}

In Sec.~\ref{sec_cRPA}, we have discussed that as far as the three-orbital model is concerned, the screening effect of the Nd layer on the NiO$_2$ layer is not so appreciable. In this Appendix, we discuss this point in more detail by comparing different model including five Ni $3d$ orbitals. 

Table~\ref{param-7orb} is a list of calculated cRPA onsite interactions for the seven-orbital, five-orbital, three-orbital, and single-orbital models. 
In the seven-orbital model, the five Ni $3d$ orbitals, Nd 5\zz, and interstitial $s$ orbitals are considered. 
In the five-orbital model, only the five Ni $3d$ orbitals are taken into account. 

Here, we focus on the onsite interaction $U$ of the Ni 3\xx orbital.
In the calculation of $U$ using the cRPA, the screening processes within the orbitals included in the model are excluded in order to avoid the double counting of the screening~\cite{Aryasetiawan_2004}.
Therefore, the $U$ value is different among different models because the exclusion of the screening is done in a different way. 
 The comparison of the onsite interaction parameters shows that:
\begin{enumerate}
    \item The difference in $U$ between the seven-orbital and five-orbital models is mainly due to screening through the Nd layer 
    [type (b) and (c) in Fig.~\ref{layer-screening}], 
    which is excluded in the calculation of $U$ in the seven-orbital model.
    For example, since Ni 3\zz orbitals can hybridize with the interstitial $s$ orbital in the Nd layer, the screening channel shown in the panel (b) can be active. Then, the screening from the Nd layer becomes effective, in accord with the recent cRPA calculation by Sakakibara {\it et al.}~\cite{Sakakibara_arXiv}.
    \item The difference in $U$ between the seven-orbital and three-orbital models is mainly due to screening through the NiO$_2$ layer [Fig.~\ref{layer-screening}(a)] and partially contributed from the Nd-layer screening [Fig.~\ref{layer-screening}(b)].
    \item The difference in $U$ between the three-orbital and single-orbital models  arises from the screening process in Fig.~\ref{layer-screening}(c) only.
    It should be noted here that the DOS of the Nd-layer orbitals is small at the Fermi level [Fig.~\ref{Fig_Wannier_band}(c)], and then this screening channel can not be drastic [point ii) in Sec.~\ref{sec_cRPA}].
    In addition, the process in Fig.~\ref{layer-screening}(b) becomes almost irrelevant, because the hybridization between the Ni 3\xx orbital and the Nd 5\zz and interstitial $s$ orbitals is negligibly small [point iii) in Sec.~\ref{sec_cRPA}].
    The process in Fig.~\ref{layer-screening}(a) is also irrelevant because the screening process from the Ni 3\xx orbital itself is excluded in the cRPA 
    [point i) in Sec.~\ref{sec_cRPA}]. 
\end{enumerate}

Therefore, whereas the Nd-layer screening plays an important role in the seven-orbital model~\cite{Sakakibara_arXiv}, 
its effect is small in the three-orbital model.

\section{Three orbital model with the Nd 5$d_{xy}$ orbital }
\label{App_3orb_dxy}
\begin{figure}[htb]
\vspace{0cm}
\begin{center}
\includegraphics[width=0.48\textwidth, clip]{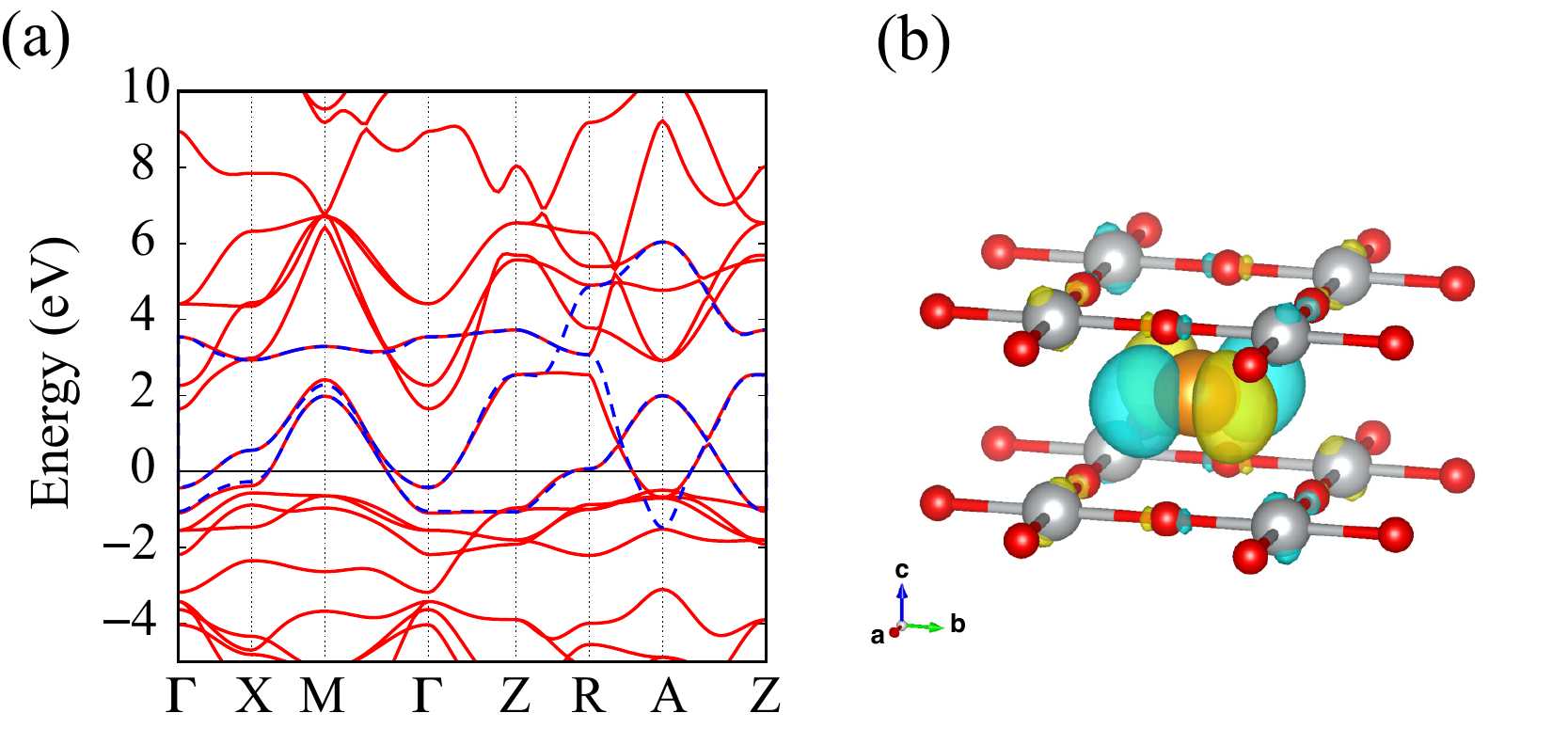}
\caption{
(a) Dispersion of three-orbital tight binding model using Ni 3\xx, Nd 5\zz, and Nd 5\xy orbitals. (blue dashed curves).  Red solid curves show the DFT band structure. 
(b) Isosurfaces (yellow: positive, light blue: negative) of constructed maximally localized Wannier function
for Nd 5\xy orbital (drawn by VESTA~\cite{Momma_2011}). 
}
\label{Fig_Wannier_sub}
\end{center}
\end{figure}

As we have already discussed in Sec.~\ref{sec_effective_model}, 
the DFT band dispersion around A point can also be reproduced by constructing the model with the Nd 5\xy orbital. 
For comparison, we show, in Fig.~\ref{Fig_Wannier_sub}, the calculated band dispersion of this alternative three-orbital tight-binding Hamiltonian [the panel (a)] and the resulting maximally localized Wannier function
for Nd 5\xy orbital [the panel (b)]. Compared with the Wannier-interpolated band in Fig.~\ref{Fig_Wannier_band}(a), we see differences in the dispersion of the conduction band. 

\bibliography{apssamp}

\providecommand{\noopsort}[1]{}\providecommand{\singleletter}[1]{#1}%
\begin{thebibliography}{73}%
\makeatletter
\providecommand \@ifxundefined [1]{%
 \@ifx{#1\undefined}
}%
\providecommand \@ifnum [1]{%
 \ifnum #1\expandafter \@firstoftwo
 \else \expandafter \@secondoftwo
 \fi
}%
\providecommand \@ifx [1]{%
 \ifx #1\expandafter \@firstoftwo
 \else \expandafter \@secondoftwo
 \fi
}%
\providecommand \natexlab [1]{#1}%
\providecommand \enquote  [1]{``#1''}%
\providecommand \bibnamefont  [1]{#1}%
\providecommand \bibfnamefont [1]{#1}%
\providecommand \citenamefont [1]{#1}%
\providecommand \href@noop [0]{\@secondoftwo}%
\providecommand \href [0]{\begingroup \@sanitize@url \@href}%
\providecommand \@href[1]{\@@startlink{#1}\@@href}%
\providecommand \@@href[1]{\endgroup#1\@@endlink}%
\providecommand \@sanitize@url [0]{\catcode `\\12\catcode `\$12\catcode
  `\&12\catcode `\#12\catcode `\^12\catcode `\_12\catcode `\%12\relax}%
\providecommand \@@startlink[1]{}%
\providecommand \@@endlink[0]{}%
\providecommand \url  [0]{\begingroup\@sanitize@url \@url }%
\providecommand \@url [1]{\endgroup\@href {#1}{\urlprefix }}%
\providecommand \urlprefix  [0]{URL }%
\providecommand \Eprint [0]{\href }%
\providecommand \doibase [0]{https://doi.org/}%
\providecommand \selectlanguage [0]{\@gobble}%
\providecommand \bibinfo  [0]{\@secondoftwo}%
\providecommand \bibfield  [0]{\@secondoftwo}%
\providecommand \translation [1]{[#1]}%
\providecommand \BibitemOpen [0]{}%
\providecommand \bibitemStop [0]{}%
\providecommand \bibitemNoStop [0]{.\EOS\space}%
\providecommand \EOS [0]{\spacefactor3000\relax}%
\providecommand \BibitemShut  [1]{\csname bibitem#1\endcsname}%
\let\auto@bib@innerbib\@empty
\bibitem [{\citenamefont {Bednorz}\ and\ \citenamefont
  {M{\"u}ller}(1986)}]{Bednorz_1986}%
  \BibitemOpen
  \bibfield  {author} {\bibinfo {author} {\bibfnamefont {J.~G.}\ \bibnamefont
  {Bednorz}}\ and\ \bibinfo {author} {\bibfnamefont {K.~A.}\ \bibnamefont
  {M{\"u}ller}},\ }\bibfield  {title} {\bibinfo {title} {{Possible high Tc
  superconductivity in the Ba-La-Cu-O system}},\ }\href
  {https://doi.org/10.1007/BF01303701} {\bibfield  {journal} {\bibinfo
  {journal} {Zeitschrift f{\"u}r Physik B Condensed Matter}\ }\textbf {\bibinfo
  {volume} {64}},\ \bibinfo {pages} {189} (\bibinfo {year} {1986})}\BibitemShut
  {NoStop}%
\bibitem [{\citenamefont {Stewart}(2017)}]{Stewart_2017}%
  \BibitemOpen
  \bibfield  {author} {\bibinfo {author} {\bibfnamefont {G.~R.}\ \bibnamefont
  {Stewart}},\ }\bibfield  {title} {\bibinfo {title} {Unconventional
  superconductivity},\ }\href {https://doi.org/10.1080/00018732.2017.1331615}
  {\bibfield  {journal} {\bibinfo  {journal} {Adv. Phys.}\ }\textbf {\bibinfo
  {volume} {66}},\ \bibinfo {pages} {75} (\bibinfo {year} {2017})}\BibitemShut
  {NoStop}%
\bibitem [{\citenamefont {Arita}\ \emph {et~al.}(1999)\citenamefont {Arita},
  \citenamefont {Kuroki},\ and\ \citenamefont {Aoki}}]{Arita_1999}%
  \BibitemOpen
  \bibfield  {author} {\bibinfo {author} {\bibfnamefont {R.}~\bibnamefont
  {Arita}}, \bibinfo {author} {\bibfnamefont {K.}~\bibnamefont {Kuroki}},\ and\
  \bibinfo {author} {\bibfnamefont {H.}~\bibnamefont {Aoki}},\ }\bibfield
  {title} {\bibinfo {title} {Spin-fluctuation exchange study of
  superconductivity in two- and three-dimensional single-band hubbard models},\
  }\href {https://doi.org/10.1103/PhysRevB.60.14585} {\bibfield  {journal}
  {\bibinfo  {journal} {Phys. Rev. B}\ }\textbf {\bibinfo {volume} {60}},\
  \bibinfo {pages} {14585} (\bibinfo {year} {1999})}\BibitemShut {NoStop}%
\bibitem [{\citenamefont {Arita}\ \emph {et~al.}(2000)\citenamefont {Arita},
  \citenamefont {Kuroki},\ and\ \citenamefont {Aoki}}]{Arita_2000}%
  \BibitemOpen
  \bibfield  {author} {\bibinfo {author} {\bibfnamefont {R.}~\bibnamefont
  {Arita}}, \bibinfo {author} {\bibfnamefont {K.}~\bibnamefont {Kuroki}},\ and\
  \bibinfo {author} {\bibfnamefont {H.}~\bibnamefont {Aoki}},\ }\bibfield
  {title} {\bibinfo {title} {{d- and p-Wave Superconductivity Mediated by Spin
  Fluctuations in Two- and Three-Dimensional Single-Band Repulsive Hubbard
  Model}},\ }\href {https://doi.org/10.1143/JPSJ.69.1181} {\bibfield  {journal}
  {\bibinfo  {journal} {J. Phys. Soc. Jpn.}\ }\textbf {\bibinfo {volume}
  {69}},\ \bibinfo {pages} {1181} (\bibinfo {year} {2000})}\BibitemShut
  {NoStop}%
\bibitem [{\citenamefont {Monthoux}\ and\ \citenamefont
  {Lonzarich}(1999)}]{Monthoux_1999}%
  \BibitemOpen
  \bibfield  {author} {\bibinfo {author} {\bibfnamefont {P.}~\bibnamefont
  {Monthoux}}\ and\ \bibinfo {author} {\bibfnamefont {G.~G.}\ \bibnamefont
  {Lonzarich}},\ }\bibfield  {title} {\bibinfo {title} {$\mathit{p}$-wave and
  d-wave superconductivity in quasi-two-dimensional metals},\ }\href
  {https://doi.org/10.1103/PhysRevB.59.14598} {\bibfield  {journal} {\bibinfo
  {journal} {Phys. Rev. B}\ }\textbf {\bibinfo {volume} {59}},\ \bibinfo
  {pages} {14598} (\bibinfo {year} {1999})}\BibitemShut {NoStop}%
\bibitem [{\citenamefont {Monthoux}\ and\ \citenamefont
  {Lonzarich}(2001)}]{Monthoux_2001}%
  \BibitemOpen
  \bibfield  {author} {\bibinfo {author} {\bibfnamefont {P.}~\bibnamefont
  {Monthoux}}\ and\ \bibinfo {author} {\bibfnamefont {G.~G.}\ \bibnamefont
  {Lonzarich}},\ }\bibfield  {title} {\bibinfo {title} {Magnetically mediated
  superconductivity in quasi-two and three dimensions},\ }\href
  {https://doi.org/10.1103/PhysRevB.63.054529} {\bibfield  {journal} {\bibinfo
  {journal} {Phys. Rev. B}\ }\textbf {\bibinfo {volume} {63}},\ \bibinfo
  {pages} {054529} (\bibinfo {year} {2001})}\BibitemShut {NoStop}%
\bibitem [{\citenamefont {Sakakibara}\ \emph {et~al.}(2010)\citenamefont
  {Sakakibara}, \citenamefont {Usui}, \citenamefont {Kuroki}, \citenamefont
  {Arita},\ and\ \citenamefont {Aoki}}]{Sakakibara_2010}%
  \BibitemOpen
  \bibfield  {author} {\bibinfo {author} {\bibfnamefont {H.}~\bibnamefont
  {Sakakibara}}, \bibinfo {author} {\bibfnamefont {H.}~\bibnamefont {Usui}},
  \bibinfo {author} {\bibfnamefont {K.}~\bibnamefont {Kuroki}}, \bibinfo
  {author} {\bibfnamefont {R.}~\bibnamefont {Arita}},\ and\ \bibinfo {author}
  {\bibfnamefont {H.}~\bibnamefont {Aoki}},\ }\bibfield  {title} {\bibinfo
  {title} {{Two-Orbital Model Explains the Higher Transition Temperature of the
  Single-Layer Hg-Cuprate Superconductor Compared to That of the La-Cuprate
  Superconductor}},\ }\href {https://doi.org/10.1103/PhysRevLett.105.057003}
  {\bibfield  {journal} {\bibinfo  {journal} {Phys. Rev. Lett.}\ }\textbf
  {\bibinfo {volume} {105}},\ \bibinfo {pages} {057003} (\bibinfo {year}
  {2010})}\BibitemShut {NoStop}%
\bibitem [{\citenamefont {Sakakibara}\ \emph
  {et~al.}(2012{\natexlab{a}})\citenamefont {Sakakibara}, \citenamefont
  {Suzuki}, \citenamefont {Usui}, \citenamefont {Kuroki}, \citenamefont
  {Arita}, \citenamefont {Scalapino},\ and\ \citenamefont
  {Aoki}}]{Sakakibara_2012}%
  \BibitemOpen
  \bibfield  {author} {\bibinfo {author} {\bibfnamefont {H.}~\bibnamefont
  {Sakakibara}}, \bibinfo {author} {\bibfnamefont {K.}~\bibnamefont {Suzuki}},
  \bibinfo {author} {\bibfnamefont {H.}~\bibnamefont {Usui}}, \bibinfo {author}
  {\bibfnamefont {K.}~\bibnamefont {Kuroki}}, \bibinfo {author} {\bibfnamefont
  {R.}~\bibnamefont {Arita}}, \bibinfo {author} {\bibfnamefont {D.~J.}\
  \bibnamefont {Scalapino}},\ and\ \bibinfo {author} {\bibfnamefont
  {H.}~\bibnamefont {Aoki}},\ }\bibfield  {title} {\bibinfo {title}
  {Multiorbital analysis of the effects of uniaxial and hydrostatic pressure on
  ${T}_{c}$ in the single-layered cuprate superconductors},\ }\href
  {https://doi.org/10.1103/PhysRevB.86.134520} {\bibfield  {journal} {\bibinfo
  {journal} {Phys. Rev. B}\ }\textbf {\bibinfo {volume} {86}},\ \bibinfo
  {pages} {134520} (\bibinfo {year} {2012}{\natexlab{a}})}\BibitemShut
  {NoStop}%
\bibitem [{\citenamefont {Sakakibara}\ \emph
  {et~al.}(2012{\natexlab{b}})\citenamefont {Sakakibara}, \citenamefont {Usui},
  \citenamefont {Kuroki}, \citenamefont {Arita},\ and\ \citenamefont
  {Aoki}}]{Sakakibara_2012b}%
  \BibitemOpen
  \bibfield  {author} {\bibinfo {author} {\bibfnamefont {H.}~\bibnamefont
  {Sakakibara}}, \bibinfo {author} {\bibfnamefont {H.}~\bibnamefont {Usui}},
  \bibinfo {author} {\bibfnamefont {K.}~\bibnamefont {Kuroki}}, \bibinfo
  {author} {\bibfnamefont {R.}~\bibnamefont {Arita}},\ and\ \bibinfo {author}
  {\bibfnamefont {H.}~\bibnamefont {Aoki}},\ }\bibfield  {title} {\bibinfo
  {title} {Origin of the material dependence of ${T}_{c}$ in the single-layered
  cuprates},\ }\href {https://doi.org/10.1103/PhysRevB.85.064501} {\bibfield
  {journal} {\bibinfo  {journal} {Phys. Rev. B}\ }\textbf {\bibinfo {volume}
  {85}},\ \bibinfo {pages} {064501} (\bibinfo {year}
  {2012}{\natexlab{b}})}\BibitemShut {NoStop}%
\bibitem [{\citenamefont {Norman}(2016)}]{Norman_2016}%
  \BibitemOpen
  \bibfield  {author} {\bibinfo {author} {\bibfnamefont {M.~R.}\ \bibnamefont
  {Norman}},\ }\bibfield  {title} {\bibinfo {title} {Materials design for new
  superconductors},\ }\href {https://doi.org/10.1088/0034-4885/79/7/074502}
  {\bibfield  {journal} {\bibinfo  {journal} {Reports on Progress in Physics}\
  }\textbf {\bibinfo {volume} {79}},\ \bibinfo {pages} {074502} (\bibinfo
  {year} {2016})}\BibitemShut {NoStop}%
\bibitem [{\citenamefont {Pickett}\ \emph {et~al.}(1989)\citenamefont
  {Pickett}, \citenamefont {Singh}, \citenamefont {Papaconstantopoulos},
  \citenamefont {Krakauer}, \citenamefont {Cyrot},\ and\ \citenamefont
  {Cyrot-Lackmann}}]{Pickett_1989}%
  \BibitemOpen
  \bibfield  {author} {\bibinfo {author} {\bibfnamefont {W.}~\bibnamefont
  {Pickett}}, \bibinfo {author} {\bibfnamefont {D.}~\bibnamefont {Singh}},
  \bibinfo {author} {\bibfnamefont {D.}~\bibnamefont {Papaconstantopoulos}},
  \bibinfo {author} {\bibfnamefont {H.}~\bibnamefont {Krakauer}}, \bibinfo
  {author} {\bibfnamefont {M.}~\bibnamefont {Cyrot}},\ and\ \bibinfo {author}
  {\bibfnamefont {F.}~\bibnamefont {Cyrot-Lackmann}},\ }\bibfield  {title}
  {\bibinfo {title} {{Theoretical studies of Sr$_{2}$VO$_{4}$, a charge
  conjugate analog of La$_{2}$CuO$_{4}$}},\ }\href
  {https://doi.org/10.1016/0921-4534(89)90758-2} {\bibfield  {journal}
  {\bibinfo  {journal} {Physica C-superconductivity and Its Applications -
  PHYSICA C}\ }\textbf {\bibinfo {volume} {162}},\ \bibinfo {pages} {1433}
  (\bibinfo {year} {1989})}\BibitemShut {NoStop}%
\bibitem [{\citenamefont {Imai}\ \emph {et~al.}(2005)\citenamefont {Imai},
  \citenamefont {Solovyev},\ and\ \citenamefont {Imada}}]{Imai_2005}%
  \BibitemOpen
  \bibfield  {author} {\bibinfo {author} {\bibfnamefont {Y.}~\bibnamefont
  {Imai}}, \bibinfo {author} {\bibfnamefont {I.}~\bibnamefont {Solovyev}},\
  and\ \bibinfo {author} {\bibfnamefont {M.}~\bibnamefont {Imada}},\ }\bibfield
   {title} {\bibinfo {title} {{Electronic Structure of Strongly Correlated
  Systems Emerging from Combining Path-Integral Renormalization Group with the
  Density-Functional Approach}},\ }\href
  {https://doi.org/10.1103/PhysRevLett.95.176405} {\bibfield  {journal}
  {\bibinfo  {journal} {Phys. Rev. Lett.}\ }\textbf {\bibinfo {volume} {95}},\
  \bibinfo {pages} {176405} (\bibinfo {year} {2005})}\BibitemShut {NoStop}%
\bibitem [{\citenamefont {Matsuno}\ \emph {et~al.}(2005)\citenamefont
  {Matsuno}, \citenamefont {Okimoto}, \citenamefont {Kawasaki},\ and\
  \citenamefont {Tokura}}]{Matsuno_2005}%
  \BibitemOpen
  \bibfield  {author} {\bibinfo {author} {\bibfnamefont {J.}~\bibnamefont
  {Matsuno}}, \bibinfo {author} {\bibfnamefont {Y.}~\bibnamefont {Okimoto}},
  \bibinfo {author} {\bibfnamefont {M.}~\bibnamefont {Kawasaki}},\ and\
  \bibinfo {author} {\bibfnamefont {Y.}~\bibnamefont {Tokura}},\ }\bibfield
  {title} {\bibinfo {title} {{Variation of the Electronic Structure in
  Systematically Synthesized ${\mathrm{Sr}}_{2}M{\mathrm{O}}_{4}$
  ($M=\mathrm{Ti}$, V, Cr, Mn, and Co)}},\ }\href
  {https://doi.org/10.1103/PhysRevLett.95.176404} {\bibfield  {journal}
  {\bibinfo  {journal} {Phys. Rev. Lett.}\ }\textbf {\bibinfo {volume} {95}},\
  \bibinfo {pages} {176404} (\bibinfo {year} {2005})}\BibitemShut {NoStop}%
\bibitem [{\citenamefont {Arita}\ \emph {et~al.}(2007)\citenamefont {Arita},
  \citenamefont {Yamasaki}, \citenamefont {Held}, \citenamefont {Matsuno},\
  and\ \citenamefont {Kuroki}}]{Arita_2007}%
  \BibitemOpen
  \bibfield  {author} {\bibinfo {author} {\bibfnamefont {R.}~\bibnamefont
  {Arita}}, \bibinfo {author} {\bibfnamefont {A.}~\bibnamefont {Yamasaki}},
  \bibinfo {author} {\bibfnamefont {K.}~\bibnamefont {Held}}, \bibinfo {author}
  {\bibfnamefont {J.}~\bibnamefont {Matsuno}},\ and\ \bibinfo {author}
  {\bibfnamefont {K.}~\bibnamefont {Kuroki}},\ }\bibfield  {title} {\bibinfo
  {title} {{${\mathrm{Sr}}_{2}\mathrm{V}{\mathrm{O}}_{4}$ and
  ${\mathrm{Ba}}_{2}\mathrm{V}{\mathrm{O}}_{4}$ under pressure: An orbital
  switch and potential ${d}^{1}$ superconductor}},\ }\href
  {https://doi.org/10.1103/PhysRevB.75.174521} {\bibfield  {journal} {\bibinfo
  {journal} {Phys. Rev. B}\ }\textbf {\bibinfo {volume} {75}},\ \bibinfo
  {pages} {174521} (\bibinfo {year} {2007})}\BibitemShut {NoStop}%
\bibitem [{\citenamefont {Chaloupka}\ and\ \citenamefont
  {Khaliullin}(2008)}]{Chaloupka_2008}%
  \BibitemOpen
  \bibfield  {author} {\bibinfo {author} {\bibfnamefont {J.~c.~v.}\
  \bibnamefont {Chaloupka}}\ and\ \bibinfo {author} {\bibfnamefont
  {G.}~\bibnamefont {Khaliullin}},\ }\bibfield  {title} {\bibinfo {title}
  {{Orbital Order and Possible Superconductivity in
  ${\mathrm{LaNiO}}_{3}/{\mathrm{LaMO}}_{3}$ Superlattices}},\ }\href
  {https://doi.org/10.1103/PhysRevLett.100.016404} {\bibfield  {journal}
  {\bibinfo  {journal} {Phys. Rev. Lett.}\ }\textbf {\bibinfo {volume} {100}},\
  \bibinfo {pages} {016404} (\bibinfo {year} {2008})}\BibitemShut {NoStop}%
\bibitem [{\citenamefont {Hansmann}\ \emph {et~al.}(2009)\citenamefont
  {Hansmann}, \citenamefont {Yang}, \citenamefont {Toschi}, \citenamefont
  {Khaliullin}, \citenamefont {Andersen},\ and\ \citenamefont
  {Held}}]{Hansmann_2009}%
  \BibitemOpen
  \bibfield  {author} {\bibinfo {author} {\bibfnamefont {P.}~\bibnamefont
  {Hansmann}}, \bibinfo {author} {\bibfnamefont {X.}~\bibnamefont {Yang}},
  \bibinfo {author} {\bibfnamefont {A.}~\bibnamefont {Toschi}}, \bibinfo
  {author} {\bibfnamefont {G.}~\bibnamefont {Khaliullin}}, \bibinfo {author}
  {\bibfnamefont {O.~K.}\ \bibnamefont {Andersen}},\ and\ \bibinfo {author}
  {\bibfnamefont {K.}~\bibnamefont {Held}},\ }\bibfield  {title} {\bibinfo
  {title} {{Turning a Nickelate Fermi Surface into a Cupratelike One through
  Heterostructuring}},\ }\href {https://doi.org/10.1103/PhysRevLett.103.016401}
  {\bibfield  {journal} {\bibinfo  {journal} {Phys. Rev. Lett.}\ }\textbf
  {\bibinfo {volume} {103}},\ \bibinfo {pages} {016401} (\bibinfo {year}
  {2009})}\BibitemShut {NoStop}%
\bibitem [{\citenamefont {Hansmann}\ \emph {et~al.}(2010)\citenamefont
  {Hansmann}, \citenamefont {Toschi}, \citenamefont {Yang}, \citenamefont
  {Andersen},\ and\ \citenamefont {Held}}]{Hansmann_2010}%
  \BibitemOpen
  \bibfield  {author} {\bibinfo {author} {\bibfnamefont {P.}~\bibnamefont
  {Hansmann}}, \bibinfo {author} {\bibfnamefont {A.}~\bibnamefont {Toschi}},
  \bibinfo {author} {\bibfnamefont {X.}~\bibnamefont {Yang}}, \bibinfo {author}
  {\bibfnamefont {O.~K.}\ \bibnamefont {Andersen}},\ and\ \bibinfo {author}
  {\bibfnamefont {K.}~\bibnamefont {Held}},\ }\bibfield  {title} {\bibinfo
  {title} {{Electronic structure of nickelates: From two-dimensional
  heterostructures to three-dimensional bulk materials}},\ }\href
  {https://doi.org/10.1103/PhysRevB.82.235123} {\bibfield  {journal} {\bibinfo
  {journal} {Phys. Rev. B}\ }\textbf {\bibinfo {volume} {82}},\ \bibinfo
  {pages} {235123} (\bibinfo {year} {2010})}\BibitemShut {NoStop}%
\bibitem [{\citenamefont {Han}\ \emph {et~al.}(2011)\citenamefont {Han},
  \citenamefont {Wang}, \citenamefont {Marianetti},\ and\ \citenamefont
  {Millis}}]{Han_2011}%
  \BibitemOpen
  \bibfield  {author} {\bibinfo {author} {\bibfnamefont {M.~J.}\ \bibnamefont
  {Han}}, \bibinfo {author} {\bibfnamefont {X.}~\bibnamefont {Wang}}, \bibinfo
  {author} {\bibfnamefont {C.~A.}\ \bibnamefont {Marianetti}},\ and\ \bibinfo
  {author} {\bibfnamefont {A.~J.}\ \bibnamefont {Millis}},\ }\bibfield  {title}
  {\bibinfo {title} {Dynamical mean-field theory of nickelate superlattices},\
  }\href {https://doi.org/10.1103/PhysRevLett.107.206804} {\bibfield  {journal}
  {\bibinfo  {journal} {Phys. Rev. Lett.}\ }\textbf {\bibinfo {volume} {107}},\
  \bibinfo {pages} {206804} (\bibinfo {year} {2011})}\BibitemShut {NoStop}%
\bibitem [{\citenamefont {Kim}\ \emph {et~al.}(2008)\citenamefont {Kim},
  \citenamefont {Jin}, \citenamefont {Moon}, \citenamefont {Kim}, \citenamefont
  {Park}, \citenamefont {Leem}, \citenamefont {Yu}, \citenamefont {Noh},
  \citenamefont {Kim}, \citenamefont {Oh}, \citenamefont {Park}, \citenamefont
  {Durairaj}, \citenamefont {Cao},\ and\ \citenamefont {Rotenberg}}]{Kim_2008}%
  \BibitemOpen
  \bibfield  {author} {\bibinfo {author} {\bibfnamefont {B.~J.}\ \bibnamefont
  {Kim}}, \bibinfo {author} {\bibfnamefont {H.}~\bibnamefont {Jin}}, \bibinfo
  {author} {\bibfnamefont {S.~J.}\ \bibnamefont {Moon}}, \bibinfo {author}
  {\bibfnamefont {J.-Y.}\ \bibnamefont {Kim}}, \bibinfo {author} {\bibfnamefont
  {B.-G.}\ \bibnamefont {Park}}, \bibinfo {author} {\bibfnamefont {C.~S.}\
  \bibnamefont {Leem}}, \bibinfo {author} {\bibfnamefont {J.}~\bibnamefont
  {Yu}}, \bibinfo {author} {\bibfnamefont {T.~W.}\ \bibnamefont {Noh}},
  \bibinfo {author} {\bibfnamefont {C.}~\bibnamefont {Kim}}, \bibinfo {author}
  {\bibfnamefont {S.-J.}\ \bibnamefont {Oh}}, \bibinfo {author} {\bibfnamefont
  {J.-H.}\ \bibnamefont {Park}}, \bibinfo {author} {\bibfnamefont
  {V.}~\bibnamefont {Durairaj}}, \bibinfo {author} {\bibfnamefont
  {G.}~\bibnamefont {Cao}},\ and\ \bibinfo {author} {\bibfnamefont
  {E.}~\bibnamefont {Rotenberg}},\ }\bibfield  {title} {\bibinfo {title}
  {{Novel ${J}_{\mathrm{eff}}=1/2$ Mott State Induced by Relativistic
  Spin-Orbit Coupling in ${\mathrm{Sr}}_{2}{\mathrm{IrO}}_{4}$}},\ }\href
  {https://doi.org/10.1103/PhysRevLett.101.076402} {\bibfield  {journal}
  {\bibinfo  {journal} {Phys. Rev. Lett.}\ }\textbf {\bibinfo {volume} {101}},\
  \bibinfo {pages} {076402} (\bibinfo {year} {2008})}\BibitemShut {NoStop}%
\bibitem [{\citenamefont {Kim}\ \emph {et~al.}(2009)\citenamefont {Kim},
  \citenamefont {Ohsumi}, \citenamefont {Komesu}, \citenamefont {Sakai},
  \citenamefont {Morita}, \citenamefont {Takagi},\ and\ \citenamefont
  {Arima}}]{Kim_2009}%
  \BibitemOpen
  \bibfield  {author} {\bibinfo {author} {\bibfnamefont {B.~J.}\ \bibnamefont
  {Kim}}, \bibinfo {author} {\bibfnamefont {H.}~\bibnamefont {Ohsumi}},
  \bibinfo {author} {\bibfnamefont {T.}~\bibnamefont {Komesu}}, \bibinfo
  {author} {\bibfnamefont {S.}~\bibnamefont {Sakai}}, \bibinfo {author}
  {\bibfnamefont {T.}~\bibnamefont {Morita}}, \bibinfo {author} {\bibfnamefont
  {H.}~\bibnamefont {Takagi}},\ and\ \bibinfo {author} {\bibfnamefont
  {T.}~\bibnamefont {Arima}},\ }\bibfield  {title} {\bibinfo {title}
  {{Phase-Sensitive Observation of a Spin-Orbital Mott State in
  Sr$_{2}$IrO$_{4}$}},\ }\href {https://doi.org/10.1126/science.1167106}
  {\bibfield  {journal} {\bibinfo  {journal} {Science}\ }\textbf {\bibinfo
  {volume} {323}},\ \bibinfo {pages} {1329} (\bibinfo {year}
  {2009})}\BibitemShut {NoStop}%
\bibitem [{\citenamefont {Watanabe}\ \emph {et~al.}(2010)\citenamefont
  {Watanabe}, \citenamefont {Shirakawa},\ and\ \citenamefont
  {Yunoki}}]{Watanabe_2010}%
  \BibitemOpen
  \bibfield  {author} {\bibinfo {author} {\bibfnamefont {H.}~\bibnamefont
  {Watanabe}}, \bibinfo {author} {\bibfnamefont {T.}~\bibnamefont
  {Shirakawa}},\ and\ \bibinfo {author} {\bibfnamefont {S.}~\bibnamefont
  {Yunoki}},\ }\bibfield  {title} {\bibinfo {title} {{Microscopic Study of a
  Spin-Orbit-Induced Mott Insulator in Ir Oxides}},\ }\href
  {https://doi.org/10.1103/PhysRevLett.105.216410} {\bibfield  {journal}
  {\bibinfo  {journal} {Phys. Rev. Lett.}\ }\textbf {\bibinfo {volume} {105}},\
  \bibinfo {pages} {216410} (\bibinfo {year} {2010})}\BibitemShut {NoStop}%
\bibitem [{\citenamefont {Watanabe}\ \emph {et~al.}(2013)\citenamefont
  {Watanabe}, \citenamefont {Shirakawa},\ and\ \citenamefont
  {Yunoki}}]{Watanabe_2013}%
  \BibitemOpen
  \bibfield  {author} {\bibinfo {author} {\bibfnamefont {H.}~\bibnamefont
  {Watanabe}}, \bibinfo {author} {\bibfnamefont {T.}~\bibnamefont
  {Shirakawa}},\ and\ \bibinfo {author} {\bibfnamefont {S.}~\bibnamefont
  {Yunoki}},\ }\bibfield  {title} {\bibinfo {title} {{Monte Carlo Study of an
  Unconventional Superconducting Phase in Iridium Oxide
  ${J}_{\mathrm{eff}}\mathbf{=}1/2$ Mott Insulators Induced by Carrier
  Doping}},\ }\href {https://doi.org/10.1103/PhysRevLett.110.027002} {\bibfield
   {journal} {\bibinfo  {journal} {Phys. Rev. Lett.}\ }\textbf {\bibinfo
  {volume} {110}},\ \bibinfo {pages} {027002} (\bibinfo {year}
  {2013})}\BibitemShut {NoStop}%
\bibitem [{\citenamefont {Kim}\ \emph {et~al.}(2015)\citenamefont {Kim},
  \citenamefont {Sung}, \citenamefont {Denlinger},\ and\ \citenamefont
  {Kim}}]{Kim_2015}%
  \BibitemOpen
  \bibfield  {author} {\bibinfo {author} {\bibfnamefont {Y.~K.}\ \bibnamefont
  {Kim}}, \bibinfo {author} {\bibfnamefont {N.~H.}\ \bibnamefont {Sung}},
  \bibinfo {author} {\bibfnamefont {J.~D.}\ \bibnamefont {Denlinger}},\ and\
  \bibinfo {author} {\bibfnamefont {B.~J.}\ \bibnamefont {Kim}},\ }\bibfield
  {title} {\bibinfo {title} {{Observation of a d-wave gap in electron-doped
  Sr$_{2}$IrO$_{4}$}},\ }\href {https://doi.org/10.1038/nphys3503} {\bibfield
  {journal} {\bibinfo  {journal} {Nature Physics}\ }\textbf {\bibinfo {volume}
  {12}},\ \bibinfo {pages} {37 EP } (\bibinfo {year} {2015})}\BibitemShut
  {NoStop}%
\bibitem [{\citenamefont {Yan}\ \emph {et~al.}(2015)\citenamefont {Yan},
  \citenamefont {Ren}, \citenamefont {Xu}, \citenamefont {Xie}, \citenamefont
  {Tao}, \citenamefont {Choi}, \citenamefont {Lee}, \citenamefont {Choi},
  \citenamefont {Zhang},\ and\ \citenamefont {Feng}}]{Yan_2015}%
  \BibitemOpen
  \bibfield  {author} {\bibinfo {author} {\bibfnamefont {Y.~J.}\ \bibnamefont
  {Yan}}, \bibinfo {author} {\bibfnamefont {M.~Q.}\ \bibnamefont {Ren}},
  \bibinfo {author} {\bibfnamefont {H.~C.}\ \bibnamefont {Xu}}, \bibinfo
  {author} {\bibfnamefont {B.~P.}\ \bibnamefont {Xie}}, \bibinfo {author}
  {\bibfnamefont {R.}~\bibnamefont {Tao}}, \bibinfo {author} {\bibfnamefont
  {H.~Y.}\ \bibnamefont {Choi}}, \bibinfo {author} {\bibfnamefont
  {N.}~\bibnamefont {Lee}}, \bibinfo {author} {\bibfnamefont {Y.~J.}\
  \bibnamefont {Choi}}, \bibinfo {author} {\bibfnamefont {T.}~\bibnamefont
  {Zhang}},\ and\ \bibinfo {author} {\bibfnamefont {D.~L.}\ \bibnamefont
  {Feng}},\ }\bibfield  {title} {\bibinfo {title} {{Electron-Doped
  ${\mathrm{Sr}}_{2}{\mathrm{IrO}}_{4}$: An Analogue of Hole-Doped Cuprate
  Superconductors Demonstrated by Scanning Tunneling Microscopy}},\ }\href
  {https://doi.org/10.1103/PhysRevX.5.041018} {\bibfield  {journal} {\bibinfo
  {journal} {Phys. Rev. X}\ }\textbf {\bibinfo {volume} {5}},\ \bibinfo {pages}
  {041018} (\bibinfo {year} {2015})}\BibitemShut {NoStop}%
\bibitem [{\citenamefont {Arita}\ \emph {et~al.}(2012)\citenamefont {Arita},
  \citenamefont {Kune\ifmmode~\check{s}\else \v{s}\fi{}}, \citenamefont
  {Kozhevnikov}, \citenamefont {Eguiluz},\ and\ \citenamefont
  {Imada}}]{Arita_2012}%
  \BibitemOpen
  \bibfield  {author} {\bibinfo {author} {\bibfnamefont {R.}~\bibnamefont
  {Arita}}, \bibinfo {author} {\bibfnamefont {J.}~\bibnamefont
  {Kune\ifmmode~\check{s}\else \v{s}\fi{}}}, \bibinfo {author} {\bibfnamefont
  {A.~V.}\ \bibnamefont {Kozhevnikov}}, \bibinfo {author} {\bibfnamefont
  {A.~G.}\ \bibnamefont {Eguiluz}},\ and\ \bibinfo {author} {\bibfnamefont
  {M.}~\bibnamefont {Imada}},\ }\bibfield  {title} {\bibinfo {title} {{Ab
  initio Studies on the Interplay between Spin-Orbit Interaction and Coulomb
  Correlation in ${\mathrm{Sr}}_{2}{\mathrm{IrO}}_{4}$ and
  ${\mathrm{Ba}}_{2}{\mathrm{IrO}}_{4}$}},\ }\href
  {https://doi.org/10.1103/PhysRevLett.108.086403} {\bibfield  {journal}
  {\bibinfo  {journal} {Phys. Rev. Lett.}\ }\textbf {\bibinfo {volume} {108}},\
  \bibinfo {pages} {086403} (\bibinfo {year} {2012})}\BibitemShut {NoStop}%
\bibitem [{\citenamefont {Martins}\ \emph {et~al.}(2011)\citenamefont
  {Martins}, \citenamefont {Aichhorn}, \citenamefont {Vaugier},\ and\
  \citenamefont {Biermann}}]{Martins_2012}%
  \BibitemOpen
  \bibfield  {author} {\bibinfo {author} {\bibfnamefont {C.}~\bibnamefont
  {Martins}}, \bibinfo {author} {\bibfnamefont {M.}~\bibnamefont {Aichhorn}},
  \bibinfo {author} {\bibfnamefont {L.}~\bibnamefont {Vaugier}},\ and\ \bibinfo
  {author} {\bibfnamefont {S.}~\bibnamefont {Biermann}},\ }\bibfield  {title}
  {\bibinfo {title} {{Reduced Effective Spin-Orbital Degeneracy and
  Spin-Orbital Ordering in Paramagnetic Transition-Metal Oxides:
  ${\mathrm{Sr}}_{2}{\mathrm{IrO}}_{4}$ versus
  ${\mathrm{Sr}}_{2}{\mathrm{RhO}}_{4}$}},\ }\href
  {https://doi.org/10.1103/PhysRevLett.107.266404} {\bibfield  {journal}
  {\bibinfo  {journal} {Phys. Rev. Lett.}\ }\textbf {\bibinfo {volume} {107}},\
  \bibinfo {pages} {266404} (\bibinfo {year} {2011})}\BibitemShut {NoStop}%
\bibitem [{\citenamefont {Li}\ \emph {et~al.}(2019)\citenamefont {Li},
  \citenamefont {Lee}, \citenamefont {Wang}, \citenamefont {Osada},
  \citenamefont {Crossley}, \citenamefont {Lee}, \citenamefont {Cui},
  \citenamefont {Hikita},\ and\ \citenamefont {Hwang}}]{Li_2019}%
  \BibitemOpen
  \bibfield  {author} {\bibinfo {author} {\bibfnamefont {D.}~\bibnamefont
  {Li}}, \bibinfo {author} {\bibfnamefont {K.}~\bibnamefont {Lee}}, \bibinfo
  {author} {\bibfnamefont {B.~Y.}\ \bibnamefont {Wang}}, \bibinfo {author}
  {\bibfnamefont {M.}~\bibnamefont {Osada}}, \bibinfo {author} {\bibfnamefont
  {S.}~\bibnamefont {Crossley}}, \bibinfo {author} {\bibfnamefont {H.~R.}\
  \bibnamefont {Lee}}, \bibinfo {author} {\bibfnamefont {Y.}~\bibnamefont
  {Cui}}, \bibinfo {author} {\bibfnamefont {Y.}~\bibnamefont {Hikita}},\ and\
  \bibinfo {author} {\bibfnamefont {H.~Y.}\ \bibnamefont {Hwang}},\ }\bibfield
  {title} {\bibinfo {title} {Superconductivity in an infinite-layer
  nickelate},\ }\href@noop {} {\bibfield  {journal} {\bibinfo  {journal}
  {Nature}\ }\textbf {\bibinfo {volume} {572}},\ \bibinfo {pages} {624}
  (\bibinfo {year} {2019})}\BibitemShut {NoStop}%
\bibitem [{\citenamefont {Azuma}\ \emph {et~al.}(1992)\citenamefont {Azuma},
  \citenamefont {Hiroi}, \citenamefont {Takano}, \citenamefont {Bando},\ and\
  \citenamefont {Takeda}}]{Azuma_1992}%
  \BibitemOpen
  \bibfield  {author} {\bibinfo {author} {\bibfnamefont {M.}~\bibnamefont
  {Azuma}}, \bibinfo {author} {\bibfnamefont {Z.}~\bibnamefont {Hiroi}},
  \bibinfo {author} {\bibfnamefont {M.}~\bibnamefont {Takano}}, \bibinfo
  {author} {\bibfnamefont {Y.}~\bibnamefont {Bando}},\ and\ \bibinfo {author}
  {\bibfnamefont {Y.}~\bibnamefont {Takeda}},\ }\bibfield  {title} {\bibinfo
  {title} {{Superconductivity at 110 K in the infinite-layer compound
  (Sr$_{1-x}$Ca$_{x}$)$_{1-y}$CuO$_{2}$}},\ }\href
  {https://doi.org/10.1038/356775a0} {\bibfield  {journal} {\bibinfo  {journal}
  {Nature}\ }\textbf {\bibinfo {volume} {356}},\ \bibinfo {pages} {775}
  (\bibinfo {year} {1992})}\BibitemShut {NoStop}%
\bibitem [{Bot()}]{Botana_arXiv}%
  \BibitemOpen
  \href@noop {} {}\bibinfo {note} {A. S. Botana and M. R. Norman,
  arXiv:1908.10946}\BibitemShut {NoStop}%
\bibitem [{\citenamefont {Lee}\ and\ \citenamefont {Pickett}(2004)}]{Lee_2004}%
  \BibitemOpen
  \bibfield  {author} {\bibinfo {author} {\bibfnamefont {K.-W.}\ \bibnamefont
  {Lee}}\ and\ \bibinfo {author} {\bibfnamefont {W.~E.}\ \bibnamefont
  {Pickett}},\ }\bibfield  {title} {\bibinfo {title} {{Infinite-layer
  $\mathrm{La}\mathrm{Ni}{\mathrm{O}}_{2}$: ${\mathrm{Ni}}^{1+}$ is not
  ${\mathrm{Cu}}^{2+}$}},\ }\href {https://doi.org/10.1103/PhysRevB.70.165109}
  {\bibfield  {journal} {\bibinfo  {journal} {Phys. Rev. B}\ }\textbf {\bibinfo
  {volume} {70}},\ \bibinfo {pages} {165109} (\bibinfo {year}
  {2004})}\BibitemShut {NoStop}%
\bibitem [{Sak()}]{Sakakibara_arXiv}%
  \BibitemOpen
  \href@noop {} {}\bibinfo {note} {H. Sakakibara, H. Usui, K. Suzuki, T.
  Kotani, H. Aoki, and K. Kuroki, arXiv:1909.00060}\BibitemShut {NoStop}%
\bibitem [{Jia()}]{Jiang_arXiv}%
  \BibitemOpen
  \href@noop {} {}\bibinfo {note} {M. Jiang, M. Berciu, and G. A. Sawatzky,
  arXiv:1909.02557}\BibitemShut {NoStop}%
\bibitem [{Wu_()}]{Wu_arXiv}%
  \BibitemOpen
  \href@noop {} {}\bibinfo {note} {X. Wu, D. Di Sante, T. Schwemmer, W. Hanke,
  H. Y. Hwang, S. Raghu, and R. Thomale, arXiv:1909.03015}\BibitemShut
  {NoStop}%
\bibitem [{\citenamefont {Zaanen}\ \emph {et~al.}(1985)\citenamefont {Zaanen},
  \citenamefont {Sawatzky},\ and\ \citenamefont {Allen}}]{Zaanen_1985}%
  \BibitemOpen
  \bibfield  {author} {\bibinfo {author} {\bibfnamefont {J.}~\bibnamefont
  {Zaanen}}, \bibinfo {author} {\bibfnamefont {G.~A.}\ \bibnamefont
  {Sawatzky}},\ and\ \bibinfo {author} {\bibfnamefont {J.~W.}\ \bibnamefont
  {Allen}},\ }\bibfield  {title} {\bibinfo {title} {Band gaps and electronic
  structure of transition-metal compounds},\ }\href
  {https://doi.org/10.1103/PhysRevLett.55.418} {\bibfield  {journal} {\bibinfo
  {journal} {Phys. Rev. Lett.}\ }\textbf {\bibinfo {volume} {55}},\ \bibinfo
  {pages} {418} (\bibinfo {year} {1985})}\BibitemShut {NoStop}%
\bibitem [{\citenamefont {Aryasetiawan}\ \emph {et~al.}(2004)\citenamefont
  {Aryasetiawan}, \citenamefont {Imada}, \citenamefont {Georges}, \citenamefont
  {Kotliar}, \citenamefont {Biermann},\ and\ \citenamefont
  {Lichtenstein}}]{Aryasetiawan_2004}%
  \BibitemOpen
  \bibfield  {author} {\bibinfo {author} {\bibfnamefont {F.}~\bibnamefont
  {Aryasetiawan}}, \bibinfo {author} {\bibfnamefont {M.}~\bibnamefont {Imada}},
  \bibinfo {author} {\bibfnamefont {A.}~\bibnamefont {Georges}}, \bibinfo
  {author} {\bibfnamefont {G.}~\bibnamefont {Kotliar}}, \bibinfo {author}
  {\bibfnamefont {S.}~\bibnamefont {Biermann}},\ and\ \bibinfo {author}
  {\bibfnamefont {A.~I.}\ \bibnamefont {Lichtenstein}},\ }\bibfield  {title}
  {\bibinfo {title} {Frequency-dependent local interactions and low-energy
  effective models from electronic structure calculations},\ }\href
  {https://doi.org/10.1103/PhysRevB.70.195104} {\bibfield  {journal} {\bibinfo
  {journal} {Phys. Rev. B}\ }\textbf {\bibinfo {volume} {70}},\ \bibinfo
  {pages} {195104} (\bibinfo {year} {2004})}\BibitemShut {NoStop}%
\bibitem [{\citenamefont {Giannozzi}\ \emph {et~al.}(2017)\citenamefont
  {Giannozzi}, \citenamefont {Andreussi}, \citenamefont {Brumme}, \citenamefont
  {Bunau}, \citenamefont {Nardelli}, \citenamefont {Calandra}, \citenamefont
  {Car}, \citenamefont {Cavazzoni}, \citenamefont {Ceresoli}, \citenamefont
  {Cococcioni}, \citenamefont {Colonna}, \citenamefont {Carnimeo},
  \citenamefont {Corso}, \citenamefont {de~Gironcoli}, \citenamefont {Delugas},
  \citenamefont {Jr}, \citenamefont {Ferretti}, \citenamefont {Floris},
  \citenamefont {Fratesi}, \citenamefont {Fugallo}, \citenamefont {Gebauer},
  \citenamefont {Gerstmann}, \citenamefont {Giustino}, \citenamefont {Gorni},
  \citenamefont {Jia}, \citenamefont {Kawamura}, \citenamefont {Ko},
  \citenamefont {Kokalj}, \citenamefont {Küçükbenli}, \citenamefont
  {Lazzeri}, \citenamefont {Marsili}, \citenamefont {Marzari}, \citenamefont
  {Mauri}, \citenamefont {Nguyen}, \citenamefont {Nguyen}, \citenamefont {de-la
  Roza}, \citenamefont {Paulatto}, \citenamefont {Poncé}, \citenamefont
  {Rocca}, \citenamefont {Sabatini}, \citenamefont {Santra}, \citenamefont
  {Schlipf}, \citenamefont {Seitsonen}, \citenamefont {Smogunov}, \citenamefont
  {Timrov}, \citenamefont {Thonhauser}, \citenamefont {Umari}, \citenamefont
  {Vast}, \citenamefont {Wu},\ and\ \citenamefont {Baroni}}]{QE-2017}%
  \BibitemOpen
  \bibfield  {author} {\bibinfo {author} {\bibfnamefont {P.}~\bibnamefont
  {Giannozzi}}, \bibinfo {author} {\bibfnamefont {O.}~\bibnamefont
  {Andreussi}}, \bibinfo {author} {\bibfnamefont {T.}~\bibnamefont {Brumme}},
  \bibinfo {author} {\bibfnamefont {O.}~\bibnamefont {Bunau}}, \bibinfo
  {author} {\bibfnamefont {M.~B.}\ \bibnamefont {Nardelli}}, \bibinfo {author}
  {\bibfnamefont {M.}~\bibnamefont {Calandra}}, \bibinfo {author}
  {\bibfnamefont {R.}~\bibnamefont {Car}}, \bibinfo {author} {\bibfnamefont
  {C.}~\bibnamefont {Cavazzoni}}, \bibinfo {author} {\bibfnamefont
  {D.}~\bibnamefont {Ceresoli}}, \bibinfo {author} {\bibfnamefont
  {M.}~\bibnamefont {Cococcioni}}, \bibinfo {author} {\bibfnamefont
  {N.}~\bibnamefont {Colonna}}, \bibinfo {author} {\bibfnamefont
  {I.}~\bibnamefont {Carnimeo}}, \bibinfo {author} {\bibfnamefont {A.~D.}\
  \bibnamefont {Corso}}, \bibinfo {author} {\bibfnamefont {S.}~\bibnamefont
  {de~Gironcoli}}, \bibinfo {author} {\bibfnamefont {P.}~\bibnamefont
  {Delugas}}, \bibinfo {author} {\bibfnamefont {R.~A.~D.}\ \bibnamefont {Jr}},
  \bibinfo {author} {\bibfnamefont {A.}~\bibnamefont {Ferretti}}, \bibinfo
  {author} {\bibfnamefont {A.}~\bibnamefont {Floris}}, \bibinfo {author}
  {\bibfnamefont {G.}~\bibnamefont {Fratesi}}, \bibinfo {author} {\bibfnamefont
  {G.}~\bibnamefont {Fugallo}}, \bibinfo {author} {\bibfnamefont
  {R.}~\bibnamefont {Gebauer}}, \bibinfo {author} {\bibfnamefont
  {U.}~\bibnamefont {Gerstmann}}, \bibinfo {author} {\bibfnamefont
  {F.}~\bibnamefont {Giustino}}, \bibinfo {author} {\bibfnamefont
  {T.}~\bibnamefont {Gorni}}, \bibinfo {author} {\bibfnamefont
  {J.}~\bibnamefont {Jia}}, \bibinfo {author} {\bibfnamefont {M.}~\bibnamefont
  {Kawamura}}, \bibinfo {author} {\bibfnamefont {H.-Y.}\ \bibnamefont {Ko}},
  \bibinfo {author} {\bibfnamefont {A.}~\bibnamefont {Kokalj}}, \bibinfo
  {author} {\bibfnamefont {E.}~\bibnamefont {Küçükbenli}}, \bibinfo {author}
  {\bibfnamefont {M.}~\bibnamefont {Lazzeri}}, \bibinfo {author} {\bibfnamefont
  {M.}~\bibnamefont {Marsili}}, \bibinfo {author} {\bibfnamefont
  {N.}~\bibnamefont {Marzari}}, \bibinfo {author} {\bibfnamefont
  {F.}~\bibnamefont {Mauri}}, \bibinfo {author} {\bibfnamefont {N.~L.}\
  \bibnamefont {Nguyen}}, \bibinfo {author} {\bibfnamefont {H.-V.}\
  \bibnamefont {Nguyen}}, \bibinfo {author} {\bibfnamefont {A.~O.}\
  \bibnamefont {de-la Roza}}, \bibinfo {author} {\bibfnamefont
  {L.}~\bibnamefont {Paulatto}}, \bibinfo {author} {\bibfnamefont
  {S.}~\bibnamefont {Poncé}}, \bibinfo {author} {\bibfnamefont
  {D.}~\bibnamefont {Rocca}}, \bibinfo {author} {\bibfnamefont
  {R.}~\bibnamefont {Sabatini}}, \bibinfo {author} {\bibfnamefont
  {B.}~\bibnamefont {Santra}}, \bibinfo {author} {\bibfnamefont
  {M.}~\bibnamefont {Schlipf}}, \bibinfo {author} {\bibfnamefont {A.~P.}\
  \bibnamefont {Seitsonen}}, \bibinfo {author} {\bibfnamefont {A.}~\bibnamefont
  {Smogunov}}, \bibinfo {author} {\bibfnamefont {I.}~\bibnamefont {Timrov}},
  \bibinfo {author} {\bibfnamefont {T.}~\bibnamefont {Thonhauser}}, \bibinfo
  {author} {\bibfnamefont {P.}~\bibnamefont {Umari}}, \bibinfo {author}
  {\bibfnamefont {N.}~\bibnamefont {Vast}}, \bibinfo {author} {\bibfnamefont
  {X.}~\bibnamefont {Wu}},\ and\ \bibinfo {author} {\bibfnamefont
  {S.}~\bibnamefont {Baroni}},\ }\bibfield  {title} {\bibinfo {title}
  {{Advanced capabilities for materials modelling with QUANTUM ESPRESSO}},\
  }\href {http://stacks.iop.org/0953-8984/29/i=46/a=465901} {\bibfield
  {journal} {\bibinfo  {journal} {Journal of Physics: Condensed Matter}\
  }\textbf {\bibinfo {volume} {29}},\ \bibinfo {pages} {465901} (\bibinfo
  {year} {2017})}\BibitemShut {NoStop}%
\bibitem [{\citenamefont {Hayward}\ and\ \citenamefont
  {Rosseinsky}(2003)}]{Hayward_2003}%
  \BibitemOpen
  \bibfield  {author} {\bibinfo {author} {\bibfnamefont {M.}~\bibnamefont
  {Hayward}}\ and\ \bibinfo {author} {\bibfnamefont {M.}~\bibnamefont
  {Rosseinsky}},\ }\bibfield  {title} {\bibinfo {title} {{Synthesis of the
  infinite layer Ni(I) phase NdNiO$_{2+x}$ by low temperature reduction of
  NdNiO$_{3}$ with sodium hydride}},\ }\href
  {https://doi.org/https://doi.org/10.1016/S1293-2558(03)00111-0} {\bibfield
  {journal} {\bibinfo  {journal} {Solid State Sciences}\ }\textbf {\bibinfo
  {volume} {5}},\ \bibinfo {pages} {839 } (\bibinfo {year} {2003})},\ \bibinfo
  {note} {international Conference on Inorganic Materials 2002}\BibitemShut
  {NoStop}%
\bibitem [{\citenamefont {Perdew}\ \emph {et~al.}(1996)\citenamefont {Perdew},
  \citenamefont {Burke},\ and\ \citenamefont {Ernzerhof}}]{Perdew_1996}%
  \BibitemOpen
  \bibfield  {author} {\bibinfo {author} {\bibfnamefont {J.~P.}\ \bibnamefont
  {Perdew}}, \bibinfo {author} {\bibfnamefont {K.}~\bibnamefont {Burke}},\ and\
  \bibinfo {author} {\bibfnamefont {M.}~\bibnamefont {Ernzerhof}},\ }\bibfield
  {title} {\bibinfo {title} {{Generalized Gradient Approximation Made
  Simple}},\ }\href {https://doi.org/10.1103/PhysRevLett.77.3865} {\bibfield
  {journal} {\bibinfo  {journal} {Phys. Rev. Lett.}\ }\textbf {\bibinfo
  {volume} {77}},\ \bibinfo {pages} {3865} (\bibinfo {year}
  {1996})}\BibitemShut {NoStop}%
\bibitem [{\citenamefont {Hamann}(2013)}]{Hamann_2013}%
  \BibitemOpen
  \bibfield  {author} {\bibinfo {author} {\bibfnamefont {D.~R.}\ \bibnamefont
  {Hamann}},\ }\bibfield  {title} {\bibinfo {title} {Optimized norm-conserving
  vanderbilt pseudopotentials},\ }\href
  {https://doi.org/10.1103/PhysRevB.88.085117} {\bibfield  {journal} {\bibinfo
  {journal} {Phys. Rev. B}\ }\textbf {\bibinfo {volume} {88}},\ \bibinfo
  {pages} {085117} (\bibinfo {year} {2013})}\BibitemShut {NoStop}%
\bibitem [{\citenamefont {van Setten}\ \emph {et~al.}(2018)\citenamefont {van
  Setten}, \citenamefont {Giantomassi}, \citenamefont {Bousquet}, \citenamefont
  {Verstraete}, \citenamefont {Hamann}, \citenamefont {Gonze},\ and\
  \citenamefont {Rignanese}}]{Setten_2018}%
  \BibitemOpen
  \bibfield  {author} {\bibinfo {author} {\bibfnamefont {M.}~\bibnamefont {van
  Setten}}, \bibinfo {author} {\bibfnamefont {M.}~\bibnamefont {Giantomassi}},
  \bibinfo {author} {\bibfnamefont {E.}~\bibnamefont {Bousquet}}, \bibinfo
  {author} {\bibfnamefont {M.}~\bibnamefont {Verstraete}}, \bibinfo {author}
  {\bibfnamefont {D.}~\bibnamefont {Hamann}}, \bibinfo {author} {\bibfnamefont
  {X.}~\bibnamefont {Gonze}},\ and\ \bibinfo {author} {\bibfnamefont {G.-M.}\
  \bibnamefont {Rignanese}},\ }\bibfield  {title} {\bibinfo {title} {{The
  PseudoDojo: Training and grading a 85 element optimized norm-conserving
  pseudopotential table}},\ }\href
  {https://doi.org/https://doi.org/10.1016/j.cpc.2018.01.012} {\bibfield
  {journal} {\bibinfo  {journal} {Computer Physics Communications}\ }\textbf
  {\bibinfo {volume} {226}},\ \bibinfo {pages} {39 } (\bibinfo {year}
  {2018})}\BibitemShut {NoStop}%
\bibitem [{\citenamefont {Marzari}\ and\ \citenamefont
  {Vanderbilt}(1997)}]{Marzari_1997}%
  \BibitemOpen
  \bibfield  {author} {\bibinfo {author} {\bibfnamefont {N.}~\bibnamefont
  {Marzari}}\ and\ \bibinfo {author} {\bibfnamefont {D.}~\bibnamefont
  {Vanderbilt}},\ }\bibfield  {title} {\bibinfo {title} {Maximally localized
  generalized wannier functions for composite energy bands},\ }\href
  {https://doi.org/10.1103/PhysRevB.56.12847} {\bibfield  {journal} {\bibinfo
  {journal} {Phys. Rev. B}\ }\textbf {\bibinfo {volume} {56}},\ \bibinfo
  {pages} {12847} (\bibinfo {year} {1997})}\BibitemShut {NoStop}%
\bibitem [{\citenamefont {Souza}\ \emph {et~al.}(2001)\citenamefont {Souza},
  \citenamefont {Marzari},\ and\ \citenamefont {Vanderbilt}}]{Souza_2001}%
  \BibitemOpen
  \bibfield  {author} {\bibinfo {author} {\bibfnamefont {I.}~\bibnamefont
  {Souza}}, \bibinfo {author} {\bibfnamefont {N.}~\bibnamefont {Marzari}},\
  and\ \bibinfo {author} {\bibfnamefont {D.}~\bibnamefont {Vanderbilt}},\
  }\bibfield  {title} {\bibinfo {title} {Maximally localized wannier functions
  for entangled energy bands},\ }\href
  {https://doi.org/10.1103/PhysRevB.65.035109} {\bibfield  {journal} {\bibinfo
  {journal} {Phys. Rev. B}\ }\textbf {\bibinfo {volume} {65}},\ \bibinfo
  {pages} {035109} (\bibinfo {year} {2001})}\BibitemShut {NoStop}%
\bibitem [{RES()}]{RESPACK_URL}%
  \BibitemOpen
  \href@noop {} {}\bibinfo {note} {\!
  https://sites.google.com/view/kazuma7k6r}\BibitemShut {NoStop}%
\bibitem [{Note1()}]{Note1}%
  \BibitemOpen
  \bibinfo {note} {Algorithms and applications of RESPACK can be found in
  Refs.~\cite
  {Nakamura_2016,Nakamura_2009,Nakamura_2008,Nohara_2009,Fujiwara_2003}.}\BibitemShut
  {Stop}%
\bibitem [{\citenamefont {\ifmmode \mbox{\c{S}}\else \c{S}\fi{}a\ifmmode
  \mbox{\c{s}}\else \c{s}\fi{}\ifmmode \imath \else \i
  \fi{}o\ifmmode~\breve{g}\else \u{g}\fi{}lu}\ \emph
  {et~al.}(2011)\citenamefont {\ifmmode \mbox{\c{S}}\else \c{S}\fi{}a\ifmmode
  \mbox{\c{s}}\else \c{s}\fi{}\ifmmode \imath \else \i
  \fi{}o\ifmmode~\breve{g}\else \u{g}\fi{}lu}, \citenamefont {Friedrich},\ and\
  \citenamefont {Bl\"ugel}}]{Sasioglu_2011}%
  \BibitemOpen
  \bibfield  {author} {\bibinfo {author} {\bibfnamefont {E.}~\bibnamefont
  {\ifmmode \mbox{\c{S}}\else \c{S}\fi{}a\ifmmode \mbox{\c{s}}\else
  \c{s}\fi{}\ifmmode \imath \else \i \fi{}o\ifmmode~\breve{g}\else
  \u{g}\fi{}lu}}, \bibinfo {author} {\bibfnamefont {C.}~\bibnamefont
  {Friedrich}},\ and\ \bibinfo {author} {\bibfnamefont {S.}~\bibnamefont
  {Bl\"ugel}},\ }\bibfield  {title} {\bibinfo {title} {Effective coulomb
  interaction in transition metals from constrained random-phase
  approximation},\ }\href {https://doi.org/10.1103/PhysRevB.83.121101}
  {\bibfield  {journal} {\bibinfo  {journal} {Phys. Rev. B}\ }\textbf {\bibinfo
  {volume} {83}},\ \bibinfo {pages} {121101} (\bibinfo {year}
  {2011})}\BibitemShut {NoStop}%
\bibitem [{\citenamefont {Yamauchi}\ \emph {et~al.}(1996)\citenamefont
  {Yamauchi}, \citenamefont {Tsukada},\ and\ \citenamefont
  {Sugino}}]{Yamauchi_1996}%
  \BibitemOpen
  \bibfield  {author} {\bibinfo {author} {\bibfnamefont {J.}~\bibnamefont
  {Yamauchi}}, \bibinfo {author} {\bibfnamefont {S.}~\bibnamefont {Tsukada},
  \bibfnamefont {Y.~Watanabe}},\ and\ \bibinfo {author} {\bibfnamefont
  {O.}~\bibnamefont {Sugino}},\ }\bibfield  {title} {\bibinfo {title}
  {{First-principles study on energetics of c-BN(001) reconstructed
  surfaces}},\ }\href@noop {} {\bibfield  {journal} {\bibinfo  {journal} {Phys.
  Rev. B}\ }\textbf {\bibinfo {volume} {54}},\ \bibinfo {pages} {5586}
  (\bibinfo {year} {1996})}\BibitemShut {NoStop}%
\bibitem [{\citenamefont {Baroni}\ \emph {et~al.}(2001)\citenamefont {Baroni},
  \citenamefont {de~Gironcoli}, \citenamefont {Dal~Corso},\ and\ \citenamefont
  {Giannozzi}}]{Baroni_2001}%
  \BibitemOpen
  \bibfield  {author} {\bibinfo {author} {\bibfnamefont {S.}~\bibnamefont
  {Baroni}}, \bibinfo {author} {\bibfnamefont {S.}~\bibnamefont
  {de~Gironcoli}}, \bibinfo {author} {\bibfnamefont {A.}~\bibnamefont
  {Dal~Corso}},\ and\ \bibinfo {author} {\bibfnamefont {P.}~\bibnamefont
  {Giannozzi}},\ }\bibfield  {title} {\bibinfo {title} {Phonons and related
  crystal properties from density-functional perturbation theory},\ }\href
  {https://doi.org/10.1103/RevModPhys.73.515} {\bibfield  {journal} {\bibinfo
  {journal} {Rev. Mod. Phys.}\ }\textbf {\bibinfo {volume} {73}},\ \bibinfo
  {pages} {515} (\bibinfo {year} {2001})}\BibitemShut {NoStop}%
\bibitem [{\citenamefont {Corso}(2014)}]{DALCORSO_2014}%
  \BibitemOpen
  \bibfield  {author} {\bibinfo {author} {\bibfnamefont {A.~D.}\ \bibnamefont
  {Corso}},\ }\bibfield  {title} {\bibinfo {title} {{Pseudopotentials periodic
  table: From H to Pu}},\ }\href@noop {} {\bibfield  {journal} {\bibinfo
  {journal} {Computational Materials Science}\ }\textbf {\bibinfo {volume}
  {95}},\ \bibinfo {pages} {337} (\bibinfo {year} {2014})}\BibitemShut
  {NoStop}%
\bibitem [{\citenamefont {Kresse}\ and\ \citenamefont
  {Furthm\"uller}(1996)}]{kresse_1996}%
  \BibitemOpen
  \bibfield  {author} {\bibinfo {author} {\bibfnamefont {G.}~\bibnamefont
  {Kresse}}\ and\ \bibinfo {author} {\bibfnamefont {J.}~\bibnamefont
  {Furthm\"uller}},\ }\bibfield  {title} {\bibinfo {title} {Efficient iterative
  schemes for ab initio total-energy calculations using a plane-wave basis
  set},\ }\href {https://doi.org/10.1103/PhysRevB.54.11169} {\bibfield
  {journal} {\bibinfo  {journal} {Phys. Rev. B}\ }\textbf {\bibinfo {volume}
  {54}},\ \bibinfo {pages} {11169} (\bibinfo {year} {1996})}\BibitemShut
  {NoStop}%
\bibitem [{\citenamefont {Bl\"ochl}(1994)}]{Blochl_1994}%
  \BibitemOpen
  \bibfield  {author} {\bibinfo {author} {\bibfnamefont {P.~E.}\ \bibnamefont
  {Bl\"ochl}},\ }\bibfield  {title} {\bibinfo {title} {Projector augmented-wave
  method},\ }\href {https://doi.org/10.1103/PhysRevB.50.17953} {\bibfield
  {journal} {\bibinfo  {journal} {Phys. Rev. B}\ }\textbf {\bibinfo {volume}
  {50}},\ \bibinfo {pages} {17953} (\bibinfo {year} {1994})}\BibitemShut
  {NoStop}%
\bibitem [{\citenamefont {Kresse}\ and\ \citenamefont
  {Joubert}(1999)}]{Kresse_1999}%
  \BibitemOpen
  \bibfield  {author} {\bibinfo {author} {\bibfnamefont {G.}~\bibnamefont
  {Kresse}}\ and\ \bibinfo {author} {\bibfnamefont {D.}~\bibnamefont
  {Joubert}},\ }\bibfield  {title} {\bibinfo {title} {From ultrasoft
  pseudopotentials to the projector augmented-wave method},\ }\href
  {https://doi.org/10.1103/PhysRevB.59.1758} {\bibfield  {journal} {\bibinfo
  {journal} {Phys. Rev. B}\ }\textbf {\bibinfo {volume} {59}},\ \bibinfo
  {pages} {1758} (\bibinfo {year} {1999})}\BibitemShut {NoStop}%
\bibitem [{\citenamefont {Ozaki}(2003)}]{Ozaki_2003}%
  \BibitemOpen
  \bibfield  {author} {\bibinfo {author} {\bibfnamefont {T.}~\bibnamefont
  {Ozaki}},\ }\bibfield  {title} {\bibinfo {title} {Variationally optimized
  atomic orbitals for large-scale electronic structures},\ }\href
  {https://doi.org/10.1103/PhysRevB.67.155108} {\bibfield  {journal} {\bibinfo
  {journal} {Phys. Rev. B}\ }\textbf {\bibinfo {volume} {67}},\ \bibinfo
  {pages} {155108} (\bibinfo {year} {2003})}\BibitemShut {NoStop}%
\bibitem [{\citenamefont {Momma}\ and\ \citenamefont
  {Izumi}(2011)}]{Momma_2011}%
  \BibitemOpen
  \bibfield  {author} {\bibinfo {author} {\bibfnamefont {K.}~\bibnamefont
  {Momma}}\ and\ \bibinfo {author} {\bibfnamefont {F.}~\bibnamefont {Izumi}},\
  }\bibfield  {title} {\bibinfo {title} {{{\it VESTA3} for three-dimensional
  visualization of crystal, volumetric and morphology data}},\ }\href
  {https://doi.org/10.1107/S0021889811038970} {\bibfield  {journal} {\bibinfo
  {journal} {Journal of Applied Crystallography}\ }\textbf {\bibinfo {volume}
  {44}},\ \bibinfo {pages} {1272} (\bibinfo {year} {2011})}\BibitemShut
  {NoStop}%
\bibitem [{\citenamefont {Matsuishi}\ \emph {et~al.}(2003)\citenamefont
  {Matsuishi}, \citenamefont {Toda}, \citenamefont {Miyakawa}, \citenamefont
  {Hayashi}, \citenamefont {Kamiya}, \citenamefont {Hirano}, \citenamefont
  {Tanaka},\ and\ \citenamefont {Hosono}}]{Matsuishi03}%
  \BibitemOpen
  \bibfield  {author} {\bibinfo {author} {\bibfnamefont {S.}~\bibnamefont
  {Matsuishi}}, \bibinfo {author} {\bibfnamefont {Y.}~\bibnamefont {Toda}},
  \bibinfo {author} {\bibfnamefont {M.}~\bibnamefont {Miyakawa}}, \bibinfo
  {author} {\bibfnamefont {K.}~\bibnamefont {Hayashi}}, \bibinfo {author}
  {\bibfnamefont {T.}~\bibnamefont {Kamiya}}, \bibinfo {author} {\bibfnamefont
  {M.}~\bibnamefont {Hirano}}, \bibinfo {author} {\bibfnamefont
  {I.}~\bibnamefont {Tanaka}},\ and\ \bibinfo {author} {\bibfnamefont
  {H.}~\bibnamefont {Hosono}},\ }\bibfield  {title} {\bibinfo {title}
  {{High-Density Electron Anions in a Nanoporous Single Crystal:
  [Ca$_{24}$Al$_{28}$O$_{64}$]$^{4+}$(4e-)}},\ }\href
  {https://doi.org/10.1126/science.1083842} {\bibfield  {journal} {\bibinfo
  {journal} {Science}\ }\textbf {\bibinfo {volume} {301}},\ \bibinfo {pages}
  {626} (\bibinfo {year} {2003})}\BibitemShut {NoStop}%
\bibitem [{\citenamefont {Hirayama}\ \emph
  {et~al.}(2018{\natexlab{a}})\citenamefont {Hirayama}, \citenamefont
  {Matsuishi}, \citenamefont {Hosono},\ and\ \citenamefont
  {Murakami}}]{Hirayama18e}%
  \BibitemOpen
  \bibfield  {author} {\bibinfo {author} {\bibfnamefont {M.}~\bibnamefont
  {Hirayama}}, \bibinfo {author} {\bibfnamefont {S.}~\bibnamefont {Matsuishi}},
  \bibinfo {author} {\bibfnamefont {H.}~\bibnamefont {Hosono}},\ and\ \bibinfo
  {author} {\bibfnamefont {S.}~\bibnamefont {Murakami}},\ }\bibfield  {title}
  {\bibinfo {title} {Electrides as a new platform of topological materials},\
  }\href {https://doi.org/10.1103/PhysRevX.8.031067} {\bibfield  {journal}
  {\bibinfo  {journal} {Phys. Rev. X}\ }\textbf {\bibinfo {volume} {8}},\
  \bibinfo {pages} {031067} (\bibinfo {year} {2018}{\natexlab{a}})}\BibitemShut
  {NoStop}%
\bibitem [{\citenamefont {Bohnen}\ \emph {et~al.}(2003)\citenamefont {Bohnen},
  \citenamefont {Heid},\ and\ \citenamefont {Krauss}}]{Bohnen_2003}%
  \BibitemOpen
  \bibfield  {author} {\bibinfo {author} {\bibfnamefont {K.~P.}\ \bibnamefont
  {Bohnen}}, \bibinfo {author} {\bibfnamefont {R.}~\bibnamefont {Heid}},\ and\
  \bibinfo {author} {\bibfnamefont {M.}~\bibnamefont {Krauss}},\ }\bibfield
  {title} {\bibinfo {title} {{Phonon dispersion and electron-phonon interaction
  for YBa$_{2}$Cu$_{3}$O$_{7}$ from first-principles calculations}},\
  }\href@noop {} {\bibfield  {journal} {\bibinfo  {journal} {EPL (Europhysics
  Letters)}\ }\textbf {\bibinfo {volume} {64}},\ \bibinfo {pages} {104}
  (\bibinfo {year} {2003})}\BibitemShut {NoStop}%
\bibitem [{\citenamefont {Giustino}\ \emph {et~al.}(2008)\citenamefont
  {Giustino}, \citenamefont {Cohen},\ and\ \citenamefont
  {Louie}}]{Giustino_2008}%
  \BibitemOpen
  \bibfield  {author} {\bibinfo {author} {\bibfnamefont {F.}~\bibnamefont
  {Giustino}}, \bibinfo {author} {\bibfnamefont {M.~L.}\ \bibnamefont
  {Cohen}},\ and\ \bibinfo {author} {\bibfnamefont {S.~G.}\ \bibnamefont
  {Louie}},\ }\bibfield  {title} {\bibinfo {title} {{Small phonon contribution
  to the photoemission kink in the copper oxide superconductors}},\ }\href@noop
  {} {\bibfield  {journal} {\bibinfo  {journal} {Nature}\ }\textbf {\bibinfo
  {volume} {452}},\ \bibinfo {pages} {975} (\bibinfo {year}
  {2008})}\BibitemShut {NoStop}%
\bibitem [{\citenamefont {Allen}\ and\ \citenamefont
  {Dynes}(1975)}]{Allen_1975}%
  \BibitemOpen
  \bibfield  {author} {\bibinfo {author} {\bibfnamefont {P.~B.}\ \bibnamefont
  {Allen}}\ and\ \bibinfo {author} {\bibfnamefont {R.~C.}\ \bibnamefont
  {Dynes}},\ }\bibfield  {title} {\bibinfo {title} {Transition temperature of
  strong-coupled superconductors reanalyzed},\ }\href
  {https://doi.org/10.1103/PhysRevB.12.905} {\bibfield  {journal} {\bibinfo
  {journal} {Phys. Rev. B}\ }\textbf {\bibinfo {volume} {12}},\ \bibinfo
  {pages} {905} (\bibinfo {year} {1975})}\BibitemShut {NoStop}%
\bibitem [{\citenamefont {Imada}\ and\ \citenamefont
  {Miyake}(2010)}]{Imada_2010}%
  \BibitemOpen
  \bibfield  {author} {\bibinfo {author} {\bibfnamefont {M.}~\bibnamefont
  {Imada}}\ and\ \bibinfo {author} {\bibfnamefont {T.}~\bibnamefont {Miyake}},\
  }\bibfield  {title} {\bibinfo {title} {{Electronic Structure Calculation by
  First Principles for Strongly Correlated Electron Systems}},\ }\href
  {https://doi.org/10.1143/JPSJ.79.112001} {\bibfield  {journal} {\bibinfo
  {journal} {J. Phys. Soc. Jpn.}\ }\textbf {\bibinfo {volume} {79}},\ \bibinfo
  {pages} {112001} (\bibinfo {year} {2010})}\BibitemShut {NoStop}%
\bibitem [{\citenamefont {Kotliar}\ \emph {et~al.}(2006)\citenamefont
  {Kotliar}, \citenamefont {Savrasov}, \citenamefont {Haule}, \citenamefont
  {Oudovenko}, \citenamefont {Parcollet},\ and\ \citenamefont
  {Marianetti}}]{Kotliar_2006}%
  \BibitemOpen
  \bibfield  {author} {\bibinfo {author} {\bibfnamefont {G.}~\bibnamefont
  {Kotliar}}, \bibinfo {author} {\bibfnamefont {S.~Y.}\ \bibnamefont
  {Savrasov}}, \bibinfo {author} {\bibfnamefont {K.}~\bibnamefont {Haule}},
  \bibinfo {author} {\bibfnamefont {V.~S.}\ \bibnamefont {Oudovenko}}, \bibinfo
  {author} {\bibfnamefont {O.}~\bibnamefont {Parcollet}},\ and\ \bibinfo
  {author} {\bibfnamefont {C.~A.}\ \bibnamefont {Marianetti}},\ }\bibfield
  {title} {\bibinfo {title} {Electronic structure calculations with dynamical
  mean-field theory},\ }\href {https://doi.org/10.1103/RevModPhys.78.865}
  {\bibfield  {journal} {\bibinfo  {journal} {Rev. Mod. Phys.}\ }\textbf
  {\bibinfo {volume} {78}},\ \bibinfo {pages} {865} (\bibinfo {year}
  {2006})}\BibitemShut {NoStop}%
\bibitem [{\citenamefont {Tadano}\ \emph {et~al.}(2019)\citenamefont {Tadano},
  \citenamefont {Nomura},\ and\ \citenamefont {Imada}}]{Tadano_2019}%
  \BibitemOpen
  \bibfield  {author} {\bibinfo {author} {\bibfnamefont {T.}~\bibnamefont
  {Tadano}}, \bibinfo {author} {\bibfnamefont {Y.}~\bibnamefont {Nomura}},\
  and\ \bibinfo {author} {\bibfnamefont {M.}~\bibnamefont {Imada}},\ }\bibfield
   {title} {\bibinfo {title} {{Ab initio derivation of an effective Hamiltonian
  for the
  ${\mathrm{La}}_{2}{\mathrm{CuO}}_{4}/{\mathrm{La}}_{1.55}{\mathrm{Sr}}_{0.45}{\mathrm{CuO}}_{4}$
  heterostructure}},\ }\href {https://doi.org/10.1103/PhysRevB.99.155148}
  {\bibfield  {journal} {\bibinfo  {journal} {Phys. Rev. B}\ }\textbf {\bibinfo
  {volume} {99}},\ \bibinfo {pages} {155148} (\bibinfo {year}
  {2019})}\BibitemShut {NoStop}%
\bibitem [{\citenamefont {Hirayama}\ \emph
  {et~al.}(2018{\natexlab{b}})\citenamefont {Hirayama}, \citenamefont {Yamaji},
  \citenamefont {Misawa},\ and\ \citenamefont {Imada}}]{Hirayama_2018}%
  \BibitemOpen
  \bibfield  {author} {\bibinfo {author} {\bibfnamefont {M.}~\bibnamefont
  {Hirayama}}, \bibinfo {author} {\bibfnamefont {Y.}~\bibnamefont {Yamaji}},
  \bibinfo {author} {\bibfnamefont {T.}~\bibnamefont {Misawa}},\ and\ \bibinfo
  {author} {\bibfnamefont {M.}~\bibnamefont {Imada}},\ }\bibfield  {title}
  {\bibinfo {title} {{Ab initio effective Hamiltonians for cuprate
  superconductors}},\ }\href {https://doi.org/10.1103/PhysRevB.98.134501}
  {\bibfield  {journal} {\bibinfo  {journal} {Phys. Rev. B}\ }\textbf {\bibinfo
  {volume} {98}},\ \bibinfo {pages} {134501} (\bibinfo {year}
  {2018}{\natexlab{b}})}\BibitemShut {NoStop}%
\bibitem [{\citenamefont {Hirayama}\ \emph {et~al.}(2019)\citenamefont
  {Hirayama}, \citenamefont {Misawa}, \citenamefont {Ohgoe}, \citenamefont
  {Yamaji},\ and\ \citenamefont {Imada}}]{Hirayama_2019}%
  \BibitemOpen
  \bibfield  {author} {\bibinfo {author} {\bibfnamefont {M.}~\bibnamefont
  {Hirayama}}, \bibinfo {author} {\bibfnamefont {T.}~\bibnamefont {Misawa}},
  \bibinfo {author} {\bibfnamefont {T.}~\bibnamefont {Ohgoe}}, \bibinfo
  {author} {\bibfnamefont {Y.}~\bibnamefont {Yamaji}},\ and\ \bibinfo {author}
  {\bibfnamefont {M.}~\bibnamefont {Imada}},\ }\bibfield  {title} {\bibinfo
  {title} {{Effective Hamiltonian for cuprate superconductors derived from
  multiscale ab initio scheme with level renormalization}},\ }\href
  {https://doi.org/10.1103/PhysRevB.99.245155} {\bibfield  {journal} {\bibinfo
  {journal} {Phys. Rev. B}\ }\textbf {\bibinfo {volume} {99}},\ \bibinfo
  {pages} {245155} (\bibinfo {year} {2019})}\BibitemShut {NoStop}%
\bibitem [{\citenamefont {Werner}\ \emph {et~al.}(2015)\citenamefont {Werner},
  \citenamefont {Sakuma}, \citenamefont {Nilsson},\ and\ \citenamefont
  {Aryasetiawan}}]{Werner_2015}%
  \BibitemOpen
  \bibfield  {author} {\bibinfo {author} {\bibfnamefont {P.}~\bibnamefont
  {Werner}}, \bibinfo {author} {\bibfnamefont {R.}~\bibnamefont {Sakuma}},
  \bibinfo {author} {\bibfnamefont {F.}~\bibnamefont {Nilsson}},\ and\ \bibinfo
  {author} {\bibfnamefont {F.}~\bibnamefont {Aryasetiawan}},\ }\bibfield
  {title} {\bibinfo {title} {{Dynamical screening in
  ${\text{La}}_{2}{\text{CuO}}_{4}$}},\ }\href
  {https://doi.org/10.1103/PhysRevB.91.125142} {\bibfield  {journal} {\bibinfo
  {journal} {Phys. Rev. B}\ }\textbf {\bibinfo {volume} {91}},\ \bibinfo
  {pages} {125142} (\bibinfo {year} {2015})}\BibitemShut {NoStop}%
\bibitem [{\citenamefont {Jang}\ \emph {et~al.}(2016)\citenamefont {Jang},
  \citenamefont {Sakakibara}, \citenamefont {Kino}, \citenamefont {Kotani},
  \citenamefont {Kuroki},\ and\ \citenamefont {Han}}]{Jang_2016}%
  \BibitemOpen
  \bibfield  {author} {\bibinfo {author} {\bibfnamefont {S.~W.}\ \bibnamefont
  {Jang}}, \bibinfo {author} {\bibfnamefont {H.}~\bibnamefont {Sakakibara}},
  \bibinfo {author} {\bibfnamefont {H.}~\bibnamefont {Kino}}, \bibinfo {author}
  {\bibfnamefont {T.}~\bibnamefont {Kotani}}, \bibinfo {author} {\bibfnamefont
  {K.}~\bibnamefont {Kuroki}},\ and\ \bibinfo {author} {\bibfnamefont {M.~J.}\
  \bibnamefont {Han}},\ }\bibfield  {title} {\bibinfo {title} {{Direct
  theoretical evidence for weaker correlations in electron-doped and Hg-based
  hole-doped cuprates}},\ }\href {https://doi.org/10.1038/srep33397} {\bibfield
   {journal} {\bibinfo  {journal} {Scientific Reports}\ }\textbf {\bibinfo
  {volume} {6}},\ \bibinfo {pages} {33397 EP } (\bibinfo {year}
  {2016})}\BibitemShut {NoStop}%
\bibitem [{\citenamefont {Nilsson}\ \emph {et~al.}(2019)\citenamefont
  {Nilsson}, \citenamefont {Karlsson},\ and\ \citenamefont
  {Aryasetiawan}}]{Nilsson_2019}%
  \BibitemOpen
  \bibfield  {author} {\bibinfo {author} {\bibfnamefont {F.}~\bibnamefont
  {Nilsson}}, \bibinfo {author} {\bibfnamefont {K.}~\bibnamefont {Karlsson}},\
  and\ \bibinfo {author} {\bibfnamefont {F.}~\bibnamefont {Aryasetiawan}},\
  }\bibfield  {title} {\bibinfo {title} {{Dynamically screened Coulomb
  interaction in the parent compounds of hole-doped cuprates: Trends and
  exceptions}},\ }\href {https://doi.org/10.1103/PhysRevB.99.075135} {\bibfield
   {journal} {\bibinfo  {journal} {Phys. Rev. B}\ }\textbf {\bibinfo {volume}
  {99}},\ \bibinfo {pages} {075135} (\bibinfo {year} {2019})}\BibitemShut
  {NoStop}%
\bibitem [{\citenamefont {Pavarini}\ \emph {et~al.}(2001)\citenamefont
  {Pavarini}, \citenamefont {Dasgupta}, \citenamefont {Saha-Dasgupta},
  \citenamefont {Jepsen},\ and\ \citenamefont {Andersen}}]{Pavarini_2001}%
  \BibitemOpen
  \bibfield  {author} {\bibinfo {author} {\bibfnamefont {E.}~\bibnamefont
  {Pavarini}}, \bibinfo {author} {\bibfnamefont {I.}~\bibnamefont {Dasgupta}},
  \bibinfo {author} {\bibfnamefont {T.}~\bibnamefont {Saha-Dasgupta}}, \bibinfo
  {author} {\bibfnamefont {O.}~\bibnamefont {Jepsen}},\ and\ \bibinfo {author}
  {\bibfnamefont {O.~K.}\ \bibnamefont {Andersen}},\ }\bibfield  {title}
  {\bibinfo {title} {{Band-Structure Trend in Hole-Doped Cuprates and
  Correlation with ${\mathit{T}}_{\mathit{c}\mathrm{max}}$}},\ }\href
  {https://doi.org/10.1103/PhysRevLett.87.047003} {\bibfield  {journal}
  {\bibinfo  {journal} {Phys. Rev. Lett.}\ }\textbf {\bibinfo {volume} {87}},\
  \bibinfo {pages} {047003} (\bibinfo {year} {2001})}\BibitemShut {NoStop}%
\bibitem [{Note2()}]{Note2}%
  \BibitemOpen
  \bibinfo {note} {We note that on ${k_z} = 0$ plane, the interstitial $s$ and
  Nd 5$d_{xy}$\protect \xspace orbitals can also hybridize to Ni
  3$d_{3z^2-r^2}$\protect \xspace orbital, whereas on ${k_z} = \pi /c$ plane,
  the hybridization is zero by symmetry. This makes a difference between ${k_z}
  = 0$ and ${k_z} = \pi /c$ planes, where the former has no Fermi pocket around
  M point while the latter has the Fermi pocket around A point.}\BibitemShut
  {Stop}%
\bibitem [{\citenamefont {Nakamura}\ \emph {et~al.}(2016)\citenamefont
  {Nakamura}, \citenamefont {Nohara}, \citenamefont {Yosimoto},\ and\
  \citenamefont {Nomura}}]{Nakamura_2016}%
  \BibitemOpen
  \bibfield  {author} {\bibinfo {author} {\bibfnamefont {K.}~\bibnamefont
  {Nakamura}}, \bibinfo {author} {\bibfnamefont {Y.}~\bibnamefont {Nohara}},
  \bibinfo {author} {\bibfnamefont {Y.}~\bibnamefont {Yosimoto}},\ and\
  \bibinfo {author} {\bibfnamefont {Y.}~\bibnamefont {Nomura}},\ }\bibfield
  {title} {\bibinfo {title} {{Ab initio $GW$ plus cumulant calculation for
  isolated band systems: Application to organic conductor
  ${(\mathrm{TMTSF})}_{2}{\mathrm{PF}}_{6}$ and transition-metal oxide
  ${\mathrm{SrVO}}_{3}$}},\ }\href {https://doi.org/10.1103/PhysRevB.93.085124}
  {\bibfield  {journal} {\bibinfo  {journal} {Phys. Rev. B}\ }\textbf {\bibinfo
  {volume} {93}},\ \bibinfo {pages} {085124} (\bibinfo {year}
  {2016})}\BibitemShut {NoStop}%
\bibitem [{\citenamefont {Nakamura}\ \emph {et~al.}(2009)\citenamefont
  {Nakamura}, \citenamefont {Yoshimoto}, \citenamefont {Kosugi}, \citenamefont
  {Arita},\ and\ \citenamefont {Imada}}]{Nakamura_2009}%
  \BibitemOpen
  \bibfield  {author} {\bibinfo {author} {\bibfnamefont {K.}~\bibnamefont
  {Nakamura}}, \bibinfo {author} {\bibfnamefont {Y.}~\bibnamefont {Yoshimoto}},
  \bibinfo {author} {\bibfnamefont {T.}~\bibnamefont {Kosugi}}, \bibinfo
  {author} {\bibfnamefont {R.}~\bibnamefont {Arita}},\ and\ \bibinfo {author}
  {\bibfnamefont {M.}~\bibnamefont {Imada}},\ }\bibfield  {title} {\bibinfo
  {title} {{Ab initio Derivation of Low-Energy Model for $\kappa$-ET Type
  Organic Conductors}},\ }\href@noop {} {\bibfield  {journal} {\bibinfo
  {journal} {J. Phys. Soc. Jpn.}\ }\textbf {\bibinfo {volume} {78}},\ \bibinfo
  {pages} {083710} (\bibinfo {year} {2009})}\BibitemShut {NoStop}%
\bibitem [{\citenamefont {Nakamura}\ \emph {et~al.}(2008)\citenamefont
  {Nakamura}, \citenamefont {Arita},\ and\ \citenamefont
  {Imada}}]{Nakamura_2008}%
  \BibitemOpen
  \bibfield  {author} {\bibinfo {author} {\bibfnamefont {K.}~\bibnamefont
  {Nakamura}}, \bibinfo {author} {\bibfnamefont {R.}~\bibnamefont {Arita}},\
  and\ \bibinfo {author} {\bibfnamefont {M.}~\bibnamefont {Imada}},\ }\bibfield
   {title} {\bibinfo {title} {{Ab initio Derivation of Low-Energy Model for
  Iron-Based Superconductors LaFeAsO and LaFePO}},\ }\href@noop {} {\bibfield
  {journal} {\bibinfo  {journal} {J. Phys. Soc. Jpn.}\ }\textbf {\bibinfo
  {volume} {77}},\ \bibinfo {pages} {093711} (\bibinfo {year}
  {2008})}\BibitemShut {NoStop}%
\bibitem [{\citenamefont {Nohara}\ \emph {et~al.}(2009)\citenamefont {Nohara},
  \citenamefont {Yamamoto},\ and\ \citenamefont {Fujiwara}}]{Nohara_2009}%
  \BibitemOpen
  \bibfield  {author} {\bibinfo {author} {\bibfnamefont {Y.}~\bibnamefont
  {Nohara}}, \bibinfo {author} {\bibfnamefont {S.}~\bibnamefont {Yamamoto}},\
  and\ \bibinfo {author} {\bibfnamefont {T.}~\bibnamefont {Fujiwara}},\
  }\bibfield  {title} {\bibinfo {title} {{Electronic structure of
  perovskite-type transition metal oxides $\text{La}M{\text{O}}_{3}$
  $(M=\text{Ti}\ensuremath{\sim}\text{Cu})$ by $\text{U}+\text{GW}$
  approximation}},\ }\href {https://doi.org/10.1103/PhysRevB.79.195110}
  {\bibfield  {journal} {\bibinfo  {journal} {Phys. Rev. B}\ }\textbf {\bibinfo
  {volume} {79}},\ \bibinfo {pages} {195110} (\bibinfo {year}
  {2009})}\BibitemShut {NoStop}%
\bibitem [{\citenamefont {Fujiwara}\ \emph {et~al.}(2003)\citenamefont
  {Fujiwara}, \citenamefont {Yamamoto},\ and\ \citenamefont
  {Ishii}}]{Fujiwara_2003}%
  \BibitemOpen
  \bibfield  {author} {\bibinfo {author} {\bibfnamefont {T.}~\bibnamefont
  {Fujiwara}}, \bibinfo {author} {\bibfnamefont {S.}~\bibnamefont {Yamamoto}},\
  and\ \bibinfo {author} {\bibfnamefont {Y.}~\bibnamefont {Ishii}},\ }\bibfield
   {title} {\bibinfo {title} {{Generalization of the Iterative Perturbation
  Theory and Metal--Insulator Transition in Multi-Orbital Hubbard Bands}},\
  }\href {https://doi.org/10.1143/JPSJ.72.777} {\bibfield  {journal} {\bibinfo
  {journal} {J. Phys. Soc. Jpn.}\ }\textbf {\bibinfo {volume} {72}},\ \bibinfo
  {pages} {777} (\bibinfo {year} {2003})}\BibitemShut {NoStop}%
\end{thebibliography}%

\end{document}